\documentclass[reprint,amsmath,amssymb,aps,pra]{revtex4-1}

\usepackage{graphicx}
\usepackage{dcolumn}
\usepackage{bm}
\usepackage{hyperref}

\usepackage{xcolor}
\newcommand \David[1] {\bgroup\noindent[\textcolor{blue}{\textbf{David}: #1}]\egroup\ignorespacesafterend}
\newcommand \MZ[1] {\bgroup\noindent[\textcolor{red}{\textbf{MZ}: #1}]\egroup\ignorespacesafterend}
\newcommand{\figref}[1]{Fig.~\ref{fig:#1}}
\newcommand{\secref}[1]{Sec.~\ref{sec:#1}}
\newcommand{\Eqref}[1]{Eq.~\ref{eq:#1}}

\newcommand{\Br}{{\boldsymbol{\mathnormal r}}}

\begin{document}

\preprint{APS/123-QED}

\title{Statistical dynamics of early creep stages in disordered materials}

\author{D. F. Castellanos}
\affiliation{
    Institute of Materials Simulation, University of Erlangen-N\"urnberg, Dr.-Mack-Stra{\ss}e 77, 90762 F\"urth, Germany
}

\author{M. Zaiser}
\affiliation{
    Institute of Materials Simulation, University of Erlangen-N\"urnberg, Dr.-Mack-Stra{\ss}e 77, 90762 F\"urth, Germany
}

\date{\today}

\begin{abstract}
When materials are loaded below their short-term strength over extended periods, a slow time-dependent process known as creep deformation takes place. During creep deformation, the structural properties of a material evolve as a function of time. By means of a generic coarse-grained mesoscopic elastoplastic model which envisages deformation as a sequence of stochastically activated discrete events, we study the creep deformation of disordered materials. We find that the structural evolution of the material during creep modifies not only the average material properties but also changes the statistics of those properties. We analyze the emergence of correlations in the strain localization and deformation activity patterns, the variation of the event rate and the evolution of the inter-event time distribution. We find that the event rate follows the Omori law of aftershocks, which is the discrete counterpart of Andrade's transient creep law, and that the exponent of these laws only depends on the microstructural heterogeneity. Finally, we find during the initial stages of transient creep a transition from Poisson distributed inter-event times towards a non-trivial power law distribution.
\end{abstract}

\maketitle


\section{\label{sec:Introduction}Introduction}
It is well-known \cite{Phillips1905,Andrade1910} that when materials are loaded below their short-term strength over extended periods, a slow time-dependent deformation known as creep takes place. Understanding creep remains of general concern in view of the long-term behavior of structural materials \cite{Lee1991} and geosystems such as Earth crust faults zones \cite{Sleep1992}. Creep deformation is characterized by several deformation regimes. Upon application of an external load, an instantaneous deformation takes place which is followed by a regime of decelerating deformation rates known as transient or primary creep. Transient creep is described by the Andrade law \cite{Andrade1910,Louchet2009}, which relates deformation to the time since application of the external load as $\epsilon \sim t^m$, with $m \approx 2/3$. Nonetheless, the value of $m$ is not universal and depends on the specific material under study \cite{Louchet2009}. Alternatively, transient creep is sometimes reported to be logarithmic, $\epsilon \sim \textrm{log}(t)$ \cite{Phillips1905,Cottrell1997}. After the transient regime, if the temperature is high enough, deformation enters the stationary or linear creep regime, $\epsilon \sim t$ \cite{Louchet2009,Main2000_GJI}. 

The description of transient creep in terms of the Andrade law has traditionally been associated with a conceptualization of creep in the framework of continuum mechanics, where plastic deformation appears as a smooth and deterministic process. Nonetheless, over the last decades, it has become clear that plasticity on small scales in which individual plastic events can be resolved appears as a stochastic and intermittent phenomenon \citep{Zaiser2008,Schneider2009}. When individual deformation events can be resolved, the slowing-down of the intermittent activity is described by the Omori law. The Omori law was originally established for earthquakes \cite{Omori1894,utsu1995} even before the early observations of transient creep which led to the Andrade law \cite{Andrade1910}. Specifically, it describes the rate of earthquake aftershocks as $\dot{n} \sim t^{-p}$, where $\dot{n}$ represents the rate of events and $p$ is typically close to 1. Omori-like behavior of plastic deformamtion activity indicators has been observed in a wide diversity of materials such as, e.g., rocks \cite{Lennartz-Sassinek2014}, porous materials \cite{Baro2013}, protein gels \cite{Leocmach2014} or bulk metallic glasses \cite{McFaul2018}.

Given the underlying fluctuating nature of the plastic activity, relating the phenomenological laws of creep to specific  microstructural processes has been a major task in understanding creep deformation. In the case of crystals, this has been achieved by means of dislocation theory \cite{Cotrell2004,Miguel2002}, where the elementary deformation events correspond to the activation of dislocation segments and their motion between metastable configurations, and the structural disorder arises from the complex stochasit patterns that are formed by the dislocation system \cite{Zaiser1999,Zaiser2006}. In the case of metallic glasses, the elementary deformation events correspond to shear transformations \cite{Argon1979,Schuh2007}. A shear transformation can be understood as a local atomic rearrangement encompassing around 100 atoms which re-arrange in a shear-like fashion in order to accommodate the local shear stress. The structural disorder inherent to a metallic glass is the result of non-equilibrium features frozen in the structure after a liquid melt is quenched into the glassy state. 

Similar creep behavior is found across a wide range of materials, such as polycrystalline metals \cite{Andrade1910}, metallic glasses \cite{Krisponeit2014_NatCom,McFaul2018}, rocks \cite{Heap2011_EPSL}, paper \cite{Rosti2010_PRL} or granular matter \cite{Nguyen2011}. Such independence of microstructural details suggests a conceptualization of creep phenomena within a framework that does not rely on material-specific microstructural processes. To this end, mesoscale elastoplastic models of disordered materials have been widely employed in recent years to reproduce and analyze intermittent plastic activity from a coarse-grained perspective \cite{Nicolas2018}. In such models, the material is discretized into a grid of mesoscopic elements of size equal to or above the length scale of the elementary deformation events, with statistically distributed elemental properties in order to represent the underlying disordered microstructure. 

In our work, we consider a mesoscale elastoplastic model based on the concept of local yield threshold, which corresponds to a mesoscopic internal state variable akin to the yield stress of continuum mechanics and characterizes the local resistance to plastic deformation. Once the locally acting shear stress overcomes the yield threshold of a mesoscopic element, a localized plastic event is triggered within that element. This process aims at capturing the essentials of plastic flow at the microscale without resolving atomistic details. The concept of yield threshold is of broad applicability, irrespective of the nature of the underlying microstructure. Advantages of the concept reside in the possibility of directly measuring the yield stress probability distribution by probing mesoscopic regions in molecular dynamics \cite{Barbot2018,Patinet2016} or DEM simulations, and by the connection it makes between plasticity and interface depinning problems \cite{BudrikisNatCom,wyart2014,Ozawa2018,Talamali2011}. 

Although mesoscale elastoplastic models have successfully reproduced much of the phenomenology associated with plastic deformation of disordered materials under strain or stress driven conditions \cite{BudrikisNatCom,Budrikis2013,Sandfeld2015,Talamali2011,wyart2014,Tuszes2017}, comparatively little attention has been given to creep deformation \cite{Merabia2017,Castellanos2018,Bouttes2013,Liu2018}. In this paper, we implement a model that is closely related to the model of creep failure by \citet{Castellanos2018}, but focus on the short-time transient dynamics and the long-time stationary flow regime. The model introduces stochastic behavior by considering thermally activated events and statistically distributed microstructural properties, which allows us to link the changes in the macroscopic creep response to the structural evolution. The model is able to capture the aging of the structure as well as mechanical rejuvenation as a function of time, external stress, temperature, and microstructural disorder. We study the spatial correlations between plastic events, strain localization, event rate and inter-event waiting time distributions during transient and stationary creep regimes.

\section{\label{sec:Model}Model}

We coarse grain the microscopic details of plastic deformation events and represent the material as a 2D array of yielding elements \cite{BudrikisNatCom,Budrikis2013,Sandfeld2015,Talamali2011,wyart2014,Tuszes2017} of linear size $L$ and volume $V_{\rm el}$ centered at positions $\Br_i$. The state of an element $i$ is represented by (i) a local stress tensor $\boldsymbol{\Sigma}(\Br_i)$ which is the superposition of the stresses resulting from external boundary conditions (for creep: temporally constant applied tractions) and internal stresses resulting from the heterogeneity of the plastic strain field, (ii) its accumulated plastic strain $\boldsymbol{\epsilon}(\Br_i)$ and (iii) a local yield threshold $\hat{\Sigma}(\Br_i)$ which characterizes the internal state of an element. Plastic deformation is assumed to be governed by the yield function
\begin{equation}
\label{eq:yield_func}
\Phi =  \sqrt{(3/2)\boldsymbol{\Sigma{'}}:\boldsymbol{\Sigma{'}}} - \hat{\Sigma} = \Sigma_{\rm eq}  - \hat{\Sigma}.
\end{equation}
Here, $\Sigma{'}$ is the deviatoric part of the stress tensor and the equivalent stress $\Sigma_{\rm eq}$ is defined in such a manner that the model amounts to a stochastic generalization of J2 plasticity (for further generalizations see \citet{BudrikisNatCom}). The yield function depends on the local yield threshold and the local stress tensor, which is computed from the external boundary conditions and the plastic strain field $\boldsymbol{\epsilon}(\Br_i)$ using standard Finite Element methodology.

On microscopic scales below the scale of resolution of our model, microscopic plastic re-arrangements are assumed to result in deformation events which produce on the element scale a tensorial plastic strain increment 
\begin{equation}
\Delta \boldsymbol{\epsilon} = \boldsymbol{\hat{\epsilon}}\Delta\epsilon
\label{eq:depsilon}
\end{equation}
where $\Delta\epsilon$ is chosen such that the local equivalent stress is reduced by a factor $0 < e \le 1$. The tensor $\boldsymbol{\hat{\epsilon}}$, which gives the 'direction' of the local strain increment is in the spirit of an associated flow rule chosen to maximize energy dissipation by setting 

\begin{equation}
\hat{\epsilon}_{ij} = \partial \Phi/\partial \Sigma_{ij}. 
\label{eq:hepsilon}
\end{equation}

\subsection{Rules for local activation of deformation events}
\label{sec:rules_activation}

The plastic flow law we adopt is characterized by the idea that deformation occurs as a stochastic sequence of local events. These events are activated according to the local values $\Phi(\Br_i)$ of the yield function. We use the following rules:
\begin{itemize}
    \item An event is activated instantaneously if $\Phi(\Br_i) > 0$. We denote this process as mechanical activation.
    \item If $\Phi(\Br_i) < 0$, an event is activated with finite rate that depends on temperature $T$ according to
    \begin{equation}
    \nu(\Br_i) = \nu_{\rm el} \exp \left( -\frac{E(\boldsymbol{\Sigma})}{k_{\rm B}T} \right). 
    \label{eq:rates0}
    \end{equation}
    where $\nu_{\rm el}$ is an attempt frequency for event activation within the element volume $V_{\rm el}$. We approximate the stress dependence of the characteristic activation energy $E$ by a linear dependency on the equivalent stress, $E = E_0 - V_{\rm A} \Sigma_{\rm eq}$where $V_{\rm A}$ is an activation volume. The activation barrier goes to zero if $\Phi = 0$, hence $E_0$ relates to the activation threshold stress via $E_0 = \hat{\Sigma} V_{\rm A}$ and we can write the 
    activation rate alternatively as
    \begin{equation}
    \nu(\Br_i) = \nu_{\rm el} \exp \left(\frac{\Phi(\Br_i)}{\Sigma_T} \right)
    \label{eq:rates}
    \end{equation}
    where the parameter $\Sigma_T = k_{\rm B} T/ V_{\rm A}$ characterizes the influence of thermal fluctuations on event activation. Besides thermal activation over a barrier, this parameter might also stem from other thermally activated processes that may trigger an event, e.g. the rate of chemical attack in chemically assisted microcracking. Beyond thermal activation, the same parameter might also be interpreted in terms of an effective magnitude of stochastic stress fluctuations of non-thermal origin affecting the region of interest, as long as such fluctuations can be represented by an effective temperature. Irrespective of the physical origin of the fluctuations, we denote this activation process as thermal activation.
    \item The duration of a deformation event is assumed negligibly small.
\end{itemize} 
In the limit where the activation thresholds $\hat{\Sigma}$ are spatially uniform and the stress changes $\Delta \epsilon$ are infinitesimally small, our model reduces for $T \to 0$ (no thermal effects) to a standard ${\rm J}_2$ plasticity model. On the other hand, at low external stress, the model reduces to a viscoplastic creep model where the rate of plastic flow is given by 
$\dot{\boldsymbol{\epsilon}} =  \boldsymbol{\hat{\epsilon}} \nu_0 \Delta \epsilon \exp(\Phi/\Sigma_T)$.

\subsection{\label{sec:threshold}Distribution of yield thresholds}

Statistical heterogeneity of the material is represented by considering the local yield thresholds $\hat{\Sigma}(\Br_i)$ as random variables. Assuming the weakest-link hypothesis at the microscopic scale \cite{Alava2009_JPD}, a mesoscopic element yields when the weakest of the microscopic elements it coarse-grains yields. In this case the mesoscopic yield thresholds are expected to follow a Weibull distribution. We consider thus a Weibull probability density with a mean value of $\hat{\Sigma}_0$ and exponent $k$

\begin{equation}
\label{eq:weibull_dist}
P(\hat{\Sigma} | \lambda, k) = \frac{k}{\lambda}\left(\frac{\hat{\Sigma}}{\lambda}\right)^{k-1}\textrm{exp}\left(-\left(\frac{\hat{\Sigma}}{\lambda}\right)^{k}\right)
\end{equation}

where the scale parameter $\lambda$ is chosen to ensure that the distribution has a mean of $\hat{\Sigma}_0$ for a specific given value $k$. The value of $k$ relates to the width of the threshold distribution and consequently characterizes the disorder of the yield thresholds. Specifically, lower values of $k$ correspond to broader distributions, i.e. higher disorder. After each deformation event, the local yield threshold of the deforming element is assigned a new value from the probability distribution (6) in order to represent structural changes. For a generalization involving strain softening, see \citet{Castellanos2018}.

\subsection{Simulation protocol}

Simulations are carried out under pure shear conditions by imposing on the free surfaces of the system spatially uniform tractions giving rise to homogeneous shear stress, in the following denoted as $\Sigma$,  which is kept fixed during a simulation. Deformation proceeds as a sequence of plastic deformation avalanches. A thermally activated event initiates an avalanche in form of a cascade of mechanically triggered events happening on a short time scale which we assume to be negligible in comparison with the inter-event time between different thermally activated events. We avoid thus competition between time scales \cite{Liu2016_PRL}, which is beyond the scope of the present investigation. The thermal events are selected by the Kinetic Monte Carlo Method with the local activation rates given by  Eq. (\ref{eq:rates}); the KMC formalism yields both the location of a the thermally activated event and the inter-event time that has elapsed since the last avalanche. After event initiation, we increase the local strain at the activated site according to Eqs. (\ref{eq:depsilon},\ref{eq:hepsilon}). We then re-compute the stress field and check whether, as a consequence of stress re-distribution, the condition $\Phi(\Br_i) < 0$ is fulfilled at any site. These sites also become activated and yield, leading to further stress changes and possible activation of further elements. The ensuing avalanche proceeds adiabatically in a series of deformation steps in each of which one or more elements are activated and yield (parallel update) until the inequality $\Phi(\Br_i) < 0$ is not fulfilled anywhere \cite{Budrikis2013,BudrikisNatCom}. The avalanche size $S$ is then defined as the total number of mechanically activated events between avalanche initiation and termination. After termination of the avalanche, a next Kinetic Monte Carlo step determines the initiation site and initiation time of the next avalanche. We repeat this process until a specified end strain is reached. To obtain statistically representative results, we perform ensemble averages over many realizations of the disorder (typically $10^3$ samples for each set of parameters).

\subsection{Parameters}

In the simulations, stress is measured in units of $\hat{\Sigma}_{0}$, the strain in units of $\hat{\Sigma}_{0}/E$ ( where $E$ is Young's modulus) and time in units of  $\nu_{0}^{-1}$. The model depends on the non-dimensional coupling constant $1-e$ which controls the characteristic fraction  of the local stress that is elastically re-distributed after an event. The model relates to particular disordered systems through the values of $e$, $\hat{\Sigma}_{0}$ and $k$. Unless otherwise stated, during this work we assume the default values $e=0.95$, $L=64$, $\Sigma_T = 0.0075$ and $k=4$.

\begin{figure}
    \includegraphics[width=0.4\textwidth]{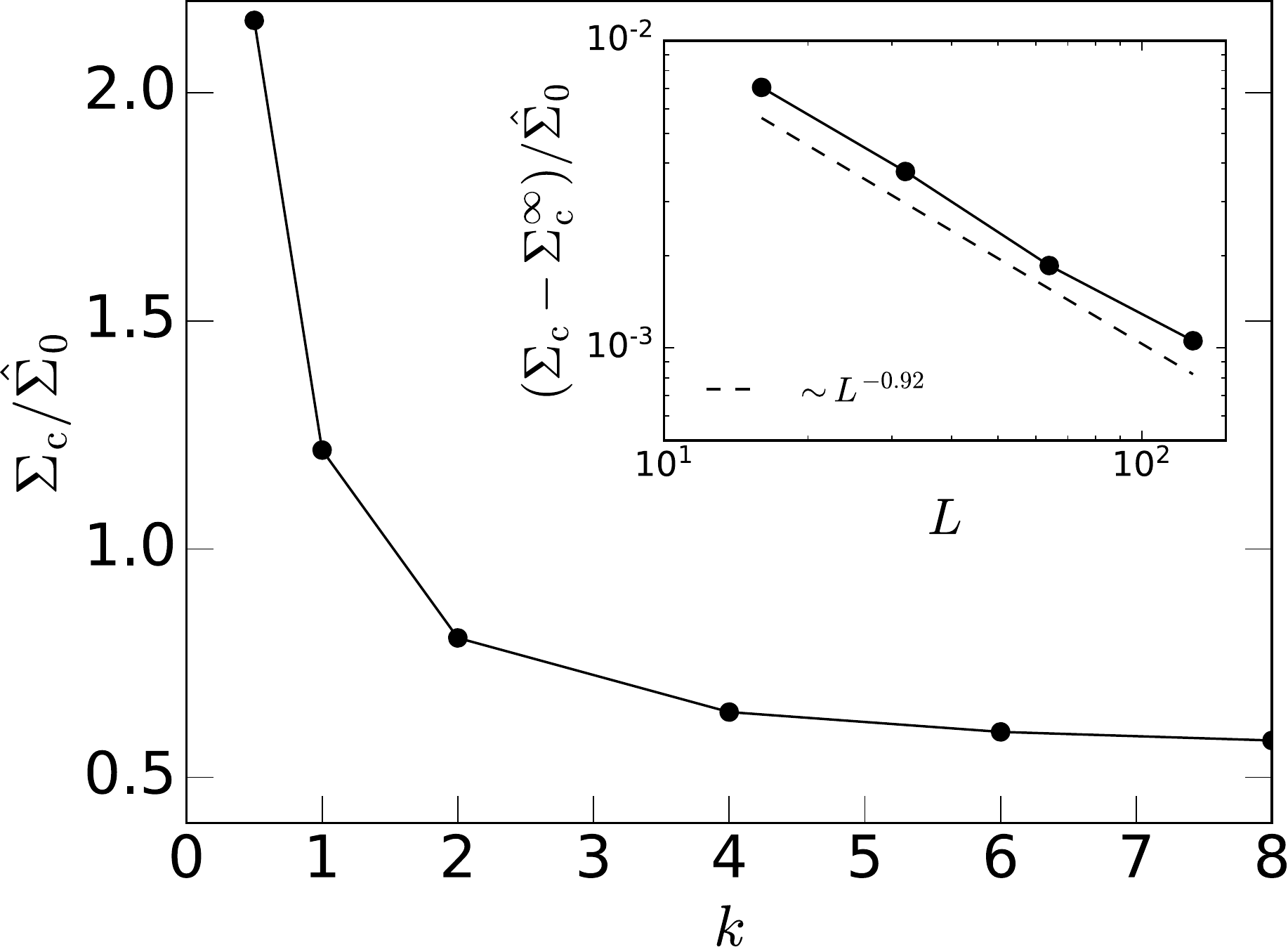}
    
    \caption{\label{fig:critical_load} Critical external stress $\Sigma_{\rm c}$ at which  macroscopic flow occurs in stress driven loading conditions, as a function of the disorder parameter $k$; inset: dependency of critical stress on system size.}
\end{figure}

Since creep loading requires the application of a constant external load, we need to determine the value of such load before performing creep tests. To this end, we use as reference the critical macroscopic yield stress $\Sigma_{\rm c}$ at which the system enters a state of sustained macroscopic plastic flow without the need for thermal activation. The value of $\Sigma_{\rm c}$ averaged over many simulations for different system parameters is shown in \figref{critical_load}. Specifically, for the default set of parameters given above, we find $\Sigma_{\rm c}=0.668$. We set a default value for the external stress of $\Sigma = 0.7 \Sigma_{\rm c}$ during our creep simulations unless otherwise stated. 

\section{Simulation results}

\subsection{Creep curves}

\begin{figure}
    \includegraphics[width=0.45\textwidth]{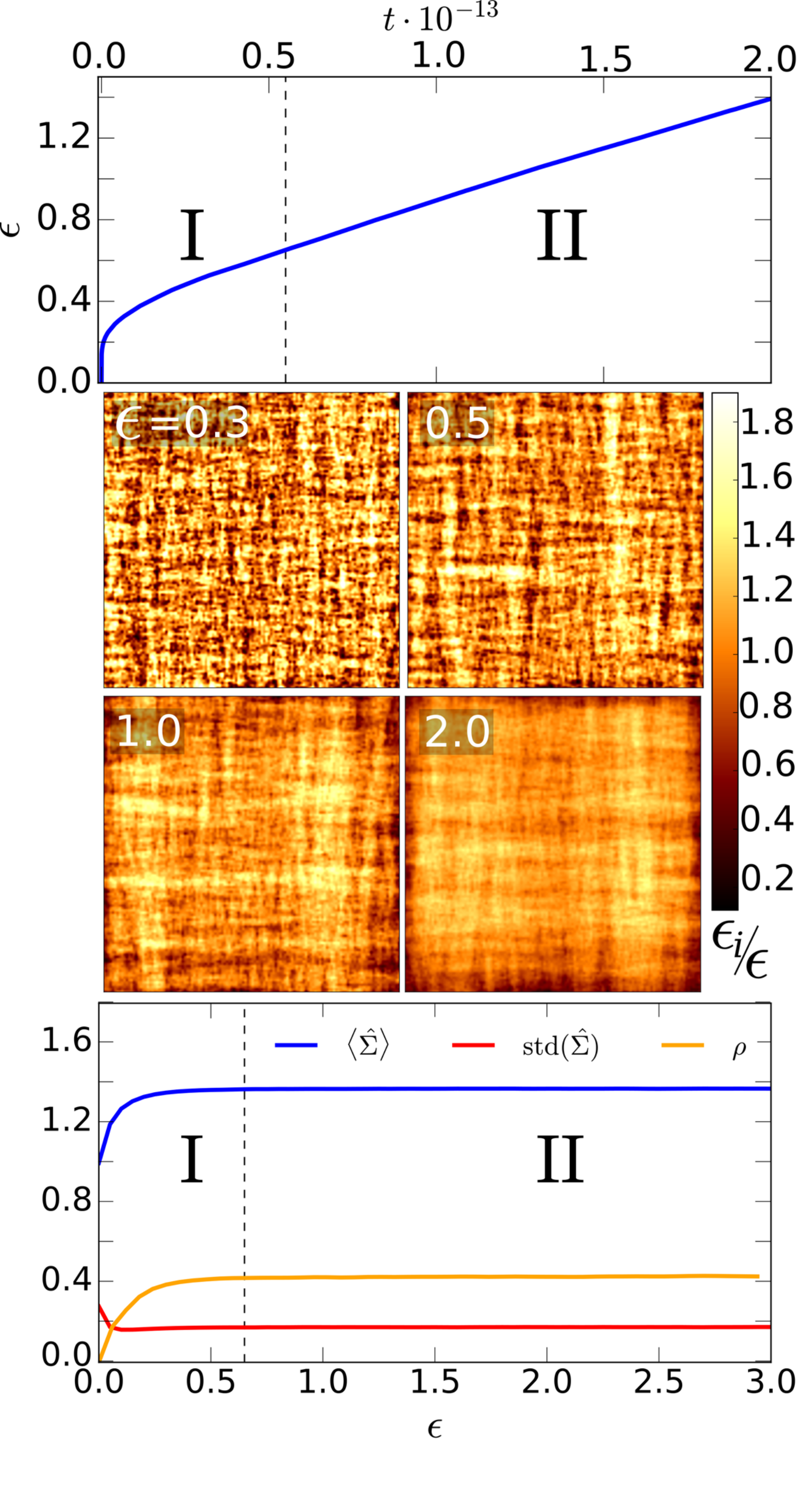}    
    \caption{Top: average creep-strain vs time curve for the default set of parameters, with $L=128$ and $\Sigma_T = 0.0075$; center: plastic strain localization patterns at different stages of the creep curve; bottom: strain evolution of the yield thresholds and the plastic event correlation coefficient.}
    \label{fig:curve_patterns} 
\end{figure}

Based on the evolution of the average strain as a function of time (see \figref{curve_patterns}, top), we observe two deformation regimes. The first regime, referred to as transient creep, is characterized by a decelerating strain rate. The second regime, referred to as stationary creep, is characterized by a constant strain rate. Tertiary creep, which is characterized by an accelerating strain rate prior to failure, is not in the focus of the present work but has been discussed in detail in a parallel study \citep{Castellanos2018}.

Immediately after the application of the load at $t=0$, all elements with a yield threshold lower than the external stress yield in an instantaneous and uncorrelated manner. Consequently, a macroscopic avalanche with a characteristic size of order $\langle S \rangle \sim L^2$ takes place. This avalanche can be interpreted as the instantaneous creep stage \cite{Louchet2009}. After the instantaneous creep, the average avalanche size quickly drops as activity is mainly due to thermal activation of individual plastic events. This gives rise to the transient regime, during which deformation continues to decelerate. After the transient regime, the deformation enters a stationary regime, where activity is controlled by thermally activated events, with an amount of mechanically triggered sequels that depends on the stress level and diverges as $\Sigma \to \Sigma_c$ but is typically small away from this critical point.

In the following, we consider different magnitudes that will help us gain insight into the internal dynamics of systems undergoing creep deformation. We study the evolution of such magnitudes along the creep process and their dependence on system parameters and deformation conditions.

\subsection{\label{sec:evolution_thresholds} Evolution of yield thresholds}

\begin{figure}
    \includegraphics[width=0.4\textwidth]{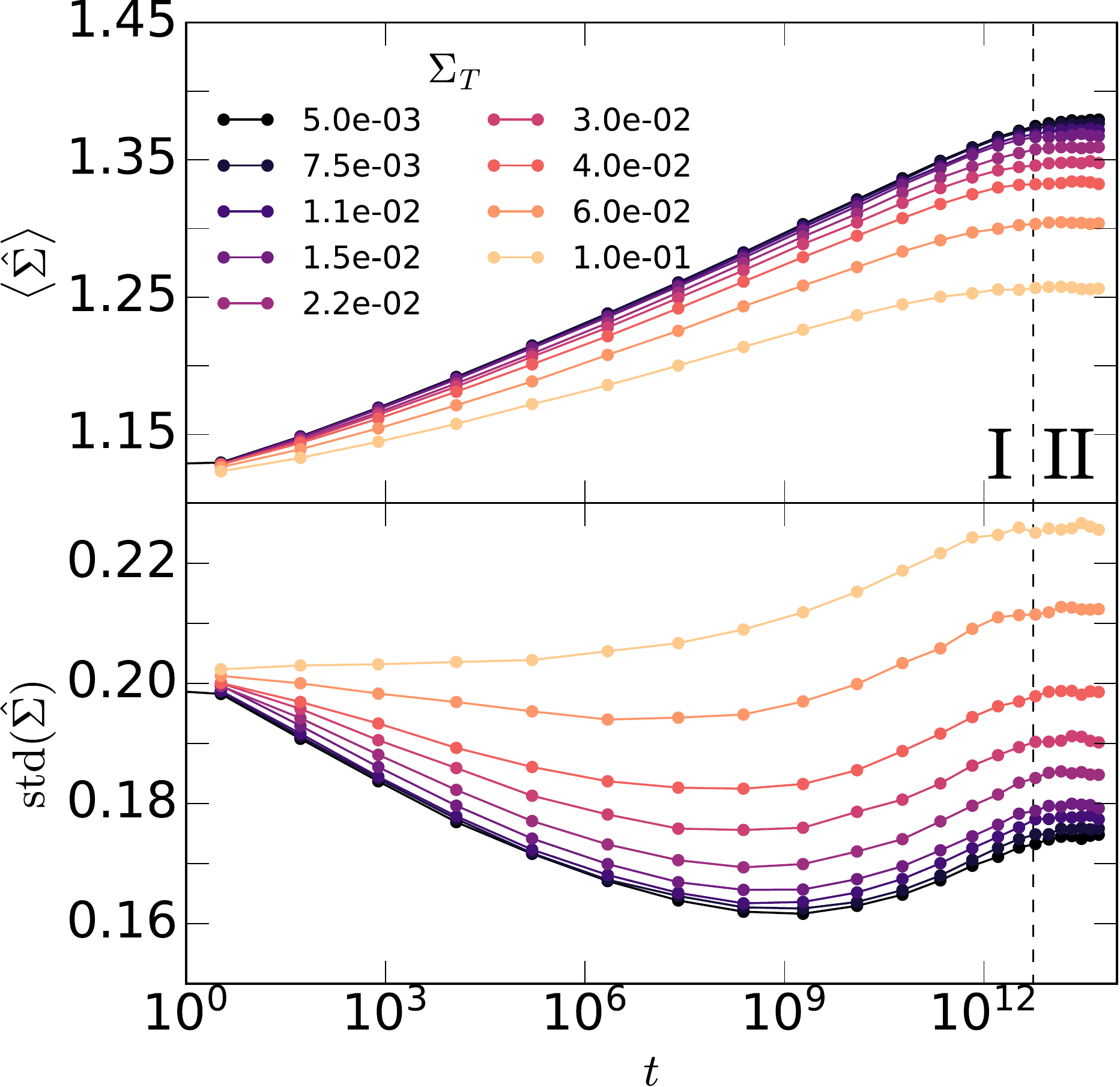}    
    \caption{Average yield threshold $\langle \hat{\Sigma} \rangle$ (top) and threshold scattering ${\rm std}(\hat{\Sigma})$ (bottom) as a function of time for different temperatures.}
    \label{fig:av_threshold_transient_temp}
\end{figure}

\begin{figure}
    \includegraphics[width=0.4\textwidth]{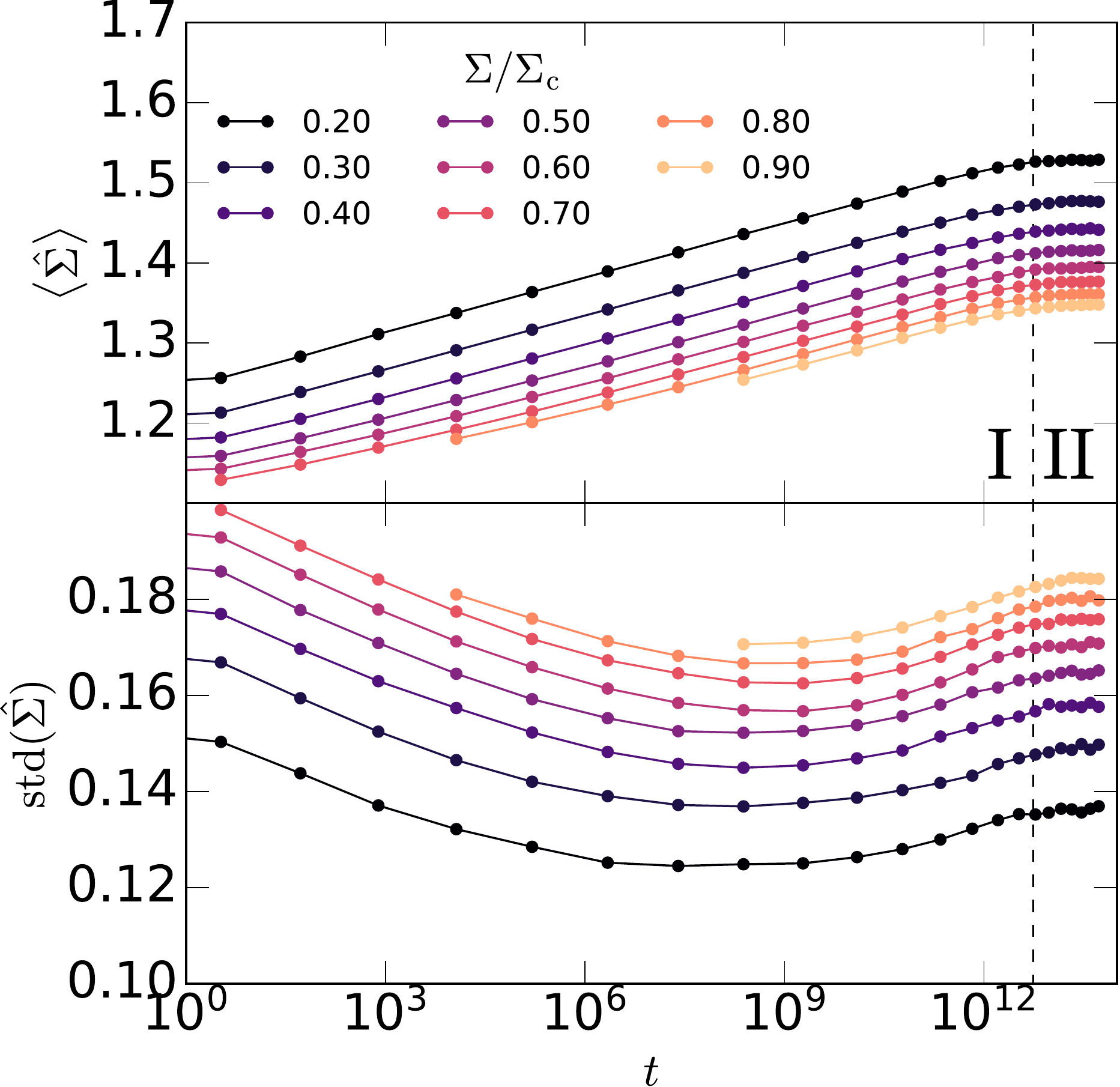}    
    \caption{Average yield threshold $\langle \hat{\Sigma} \rangle$ (top) and threshold scattering ${\rm std}(\hat{\Sigma})$ (bottom) as a function of stress for different temperatures.}
    \label{fig:av_threshold_transient_stress}
\end{figure}

We first consider the time evolution of the average local yield threshold $\langle \hat{\Sigma} \rangle$, and the threshold scatter measured by the standard deviation of thresholds ${\rm std}(\hat{\Sigma})$.
As shown in \figref{av_threshold_transient_temp} (top) and \figref{av_threshold_transient_stress} (top), the average threshold first increase from its initial value of $\langle \hat{\Sigma} \rangle=1$. This hardening is responsible for the slowing down of creep deformation in the transient regime. Commonly known as statistical hardening, it is the consequence of a survival bias due to the statistically distributed yield thresholds: The lowest threshold elements are, according to the rules of \secref{rules_activation}, more likely to yield and the re-assignment of a new threshold leads, on average, to a higher local threshold. The evolution of the average threshold can be interpreted as structural aging, by which thermally activated plastic activity leads to a more stable structure. The value of ${\rm std}(\hat{\Sigma})$ is found to drop until reaching a minimum value, after which it grows until attaining its plateau value at the stationary creep regime (see \figref{av_threshold_transient_temp} and see \figref{av_threshold_transient_stress}, bottom). The existence and location of the minimum depends on temperature (it disappears at high temperatures) but is robust upon variation of the external stress.

\begin{figure}
    \includegraphics[width=0.45\textwidth]{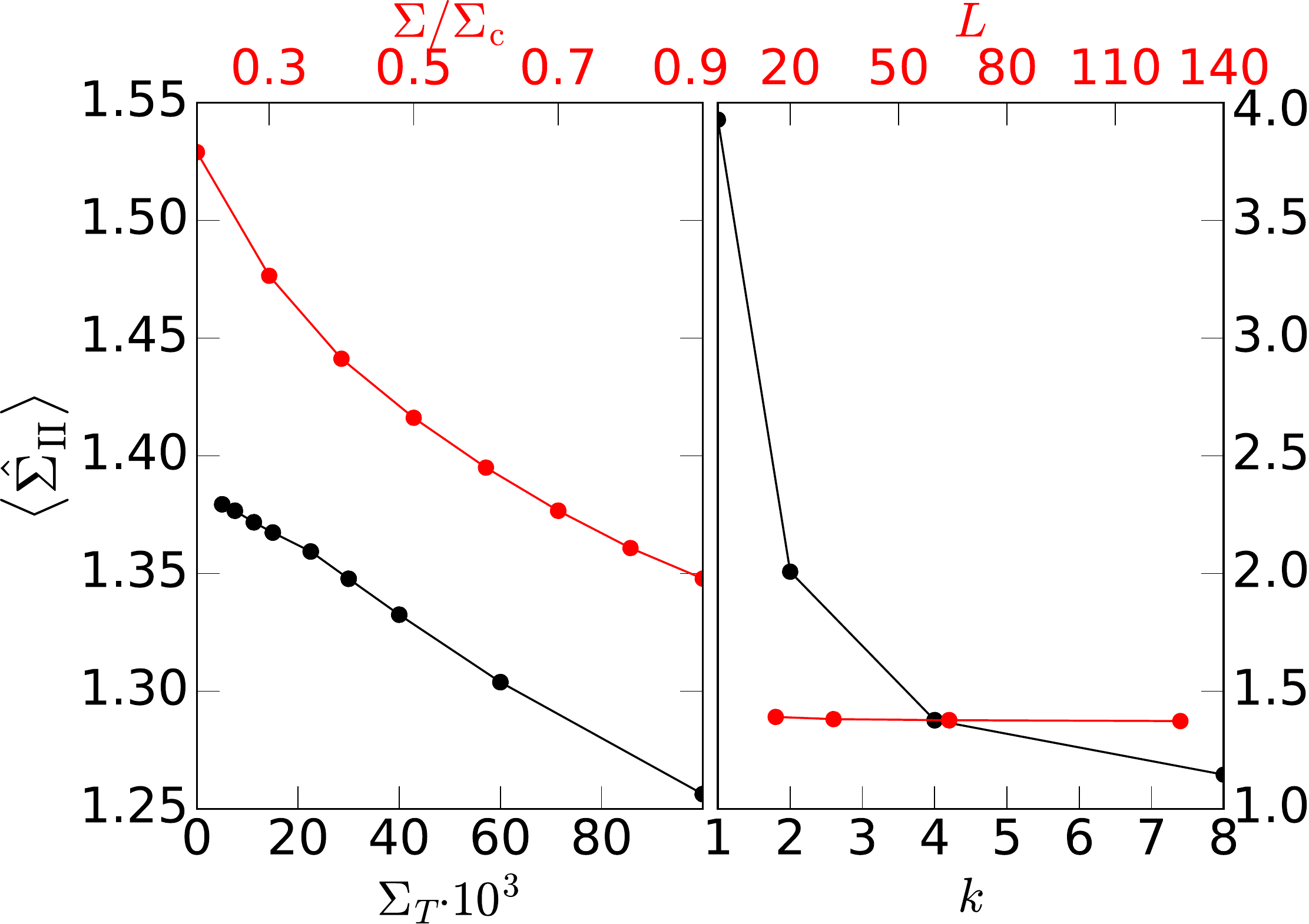}    
    \caption{Average yield threshold $\langle \hat{\Sigma}_{\rm II} \rangle$ in the stationary creep regime for different system parameters.}
    \label{fig:av_threshold_stationary}
\end{figure}

\begin{figure}
    \includegraphics[width=0.45\textwidth]{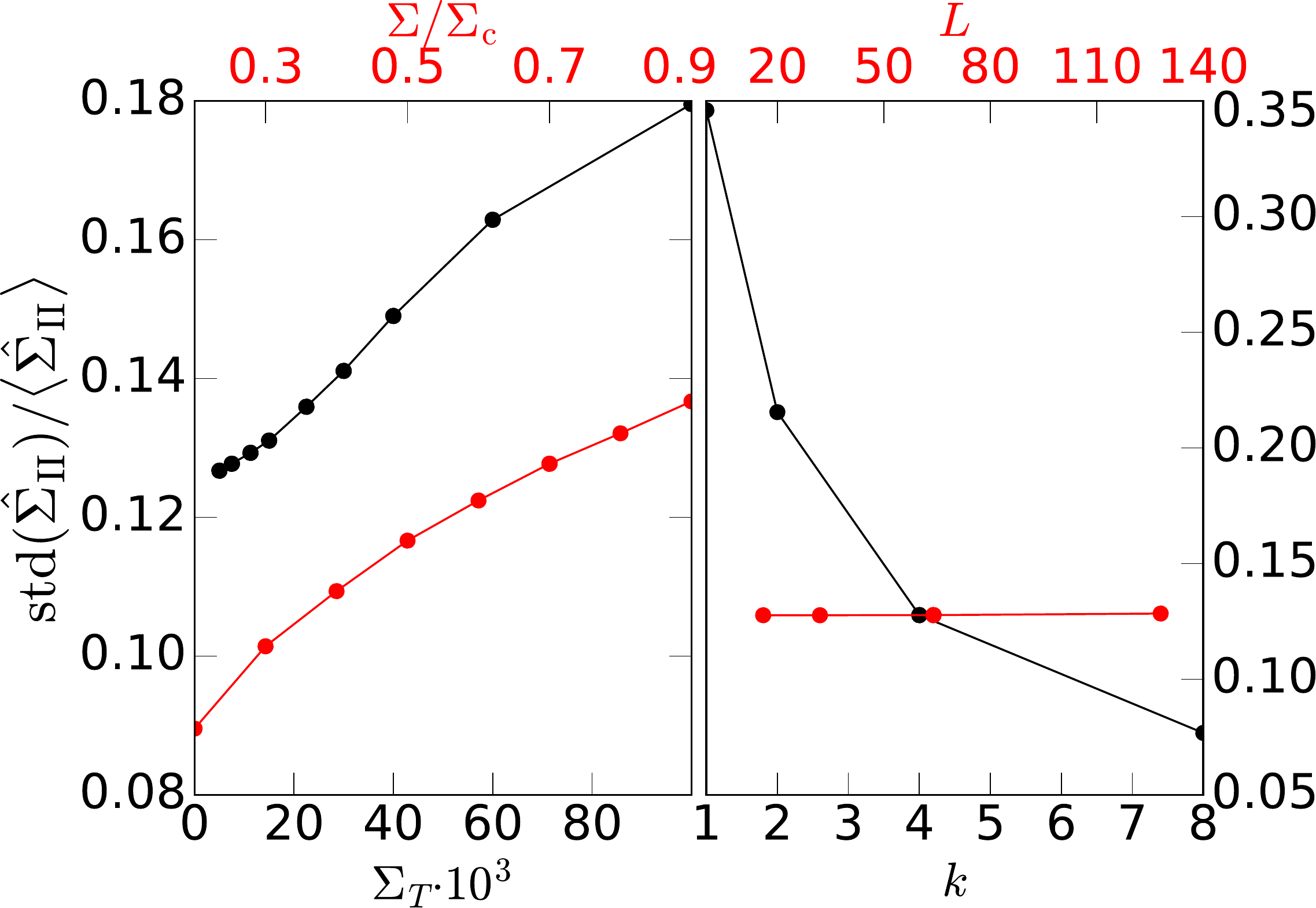}    
    \caption{Yield threshold scattering relative to the average value, ${\rm std}(\hat{\Sigma}_{\rm II})/\langle \hat{\Sigma}_{\rm II} \rangle$, in the stationary creep regime for different system parameters.}
    \label{fig:std_threshold_stationary}
\end{figure}

Statistical hardening saturates as the thresholds move to the high strength side of the distribution 
(6) until a dynamic equilibrium situation is reached where the mean and scatter of the yield thresholds in the sample no longer change. Thus, the system enters the stationary creep regime which is characterized by a constant deformation rate, as shown in \figref{curve_patterns} (bottom). 
We denote the value of the stationary average yield threshold as $\langle \hat{\Sigma}_{\rm II} \rangle$ and that of the threshold scatter as ${\rm std}(\hat{\Sigma}_{\rm II})$. \figref{av_threshold_stationary} shows the values of $\langle \hat{\Sigma}_{\rm II} \rangle$ for different system parameters, and \figref{std_threshold_stationary} shows the coefficient of variation ${\rm std}(\hat{\Sigma}_{\rm II})/\langle \hat{\Sigma}_{\rm II} \rangle$.  We find that, as the external stress is increased, the average threshold stabilizes at a lower value while the coefficient of variation increases. Therefore, an increase of the external stress reverts to some extent, the effects of aging. This stress dependence of structural stability is commonly known as mechanical rejuvenation \cite{Gruber1978,Zhang2017,Warren2008}. Increasing temperature is found to enhance the effect of mechanical rejuvenation. We find that systems with higher initial disorder (i.e., lower $k$) reach higher saturation values of the local strength. This is in line with the observations of \cite{Tuszes2017} who report, for systems loaded in displacement control, that strength increases with increasing degree of disorder. The system size is found to have no discernable influence on the statistical behavior.

\subsection{\label{sec:spatial_activity} Spatial activity patterns}

\subsubsection{Strain localization}
During the initial stages of the transient regime, internal stresses are almost absent, and the system behavior is controlled by externally applied shear stress which is spatially homogeneous. As a result, plastic deformation is activated in a spatially uniform random manner (see \figref{curve_patterns}, patterns). As deformation proceeds, the internal stress field created by previous plastic events induces elastic couplings within the system, which leads to self-organization of the deformation activity and the emergence of strain patterns in the form of non-permanent, mutually perpendicular shear bands. The orientation of the bands is the result of the loading mode: our external load induces a pure shear state with principal axes oriented along $\pm \pi/4$ with respect to the horizontal axis in Figs. (2) and (8). The plastic events follow this orientation, and, as a consequence, positive stress redistribution favoring the formation of bands takes places along the directions $0$ and $\pi/2$ with respect to the horizontal axis. However, the ensuing strain localization is only transient: As the creep strain increases during the stationary creep regime, the plastic strain pattern arising from the superposition of strain bands remains statistially homogeneous. 

\subsubsection{Event correlation}
\label{sec:event_correlation}
To quantitatively study the spatial activity, we consider the Pearson correlation coefficient $\rho$ between the locations of the triggering events of two consecutive avalanches. Specifically, we consider the projected locations over the horizontal axis, $x$, and $x^{\prime}$. As discussed by \cite{Castellanos2018}, care must be taken in the case of the symmetry breaking induced by, e.g., permanent strain localization in the form of a macroscopic shear band. However, since we study the stationary creep regime without the presence of any softening mechanism, the activity is symmetric with respect to the projection over any pair of perpendicular directions. We defined thus the coefficient

\begin{equation}
\rho(\epsilon) = \frac{ \langle x x^{\prime} \rangle_{\epsilon} - \langle x \rangle_{\epsilon}^{2} }{\sigma(x,\epsilon)^2}
\end{equation}

where $\langle \cdot \rangle_{\epsilon}$ denotes an average over a narrow strain window centered at $\epsilon$ and $\sigma(\cdot,\epsilon)$ is the standard deviation of event locations within that strain window (when no $\epsilon$ is present, the operation is performed over all the complete simulation). The evolution of the correlation coefficient $\rho$ with strain is shown in \figref{curve_patterns} (bottom). The initial lack of elastic coupling is reflected by near-zero correlation coefficient $\rho$, while on the approach to the stationary regime the coefficient saturates at a value $\rho_{\rm II}$.

\begin{figure}
    \includegraphics[width=0.44\textwidth]{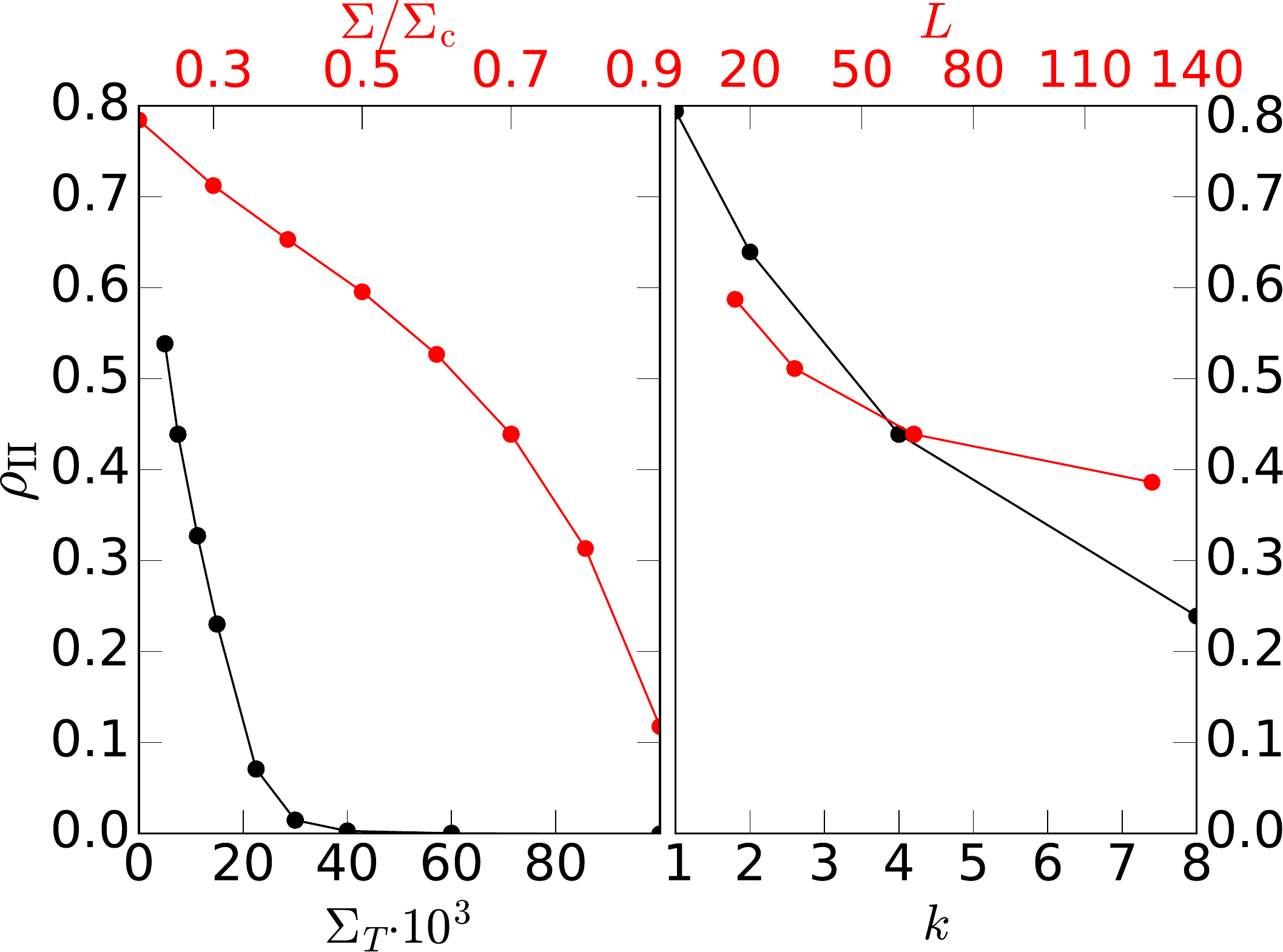}    
    \caption{Correlation coefficient during the stationary creep regime for different system parameters.}
    \label{fig:corr_coefficient_stationary}
\end{figure}

\begin{figure}
    \includegraphics[width=0.4\textwidth]{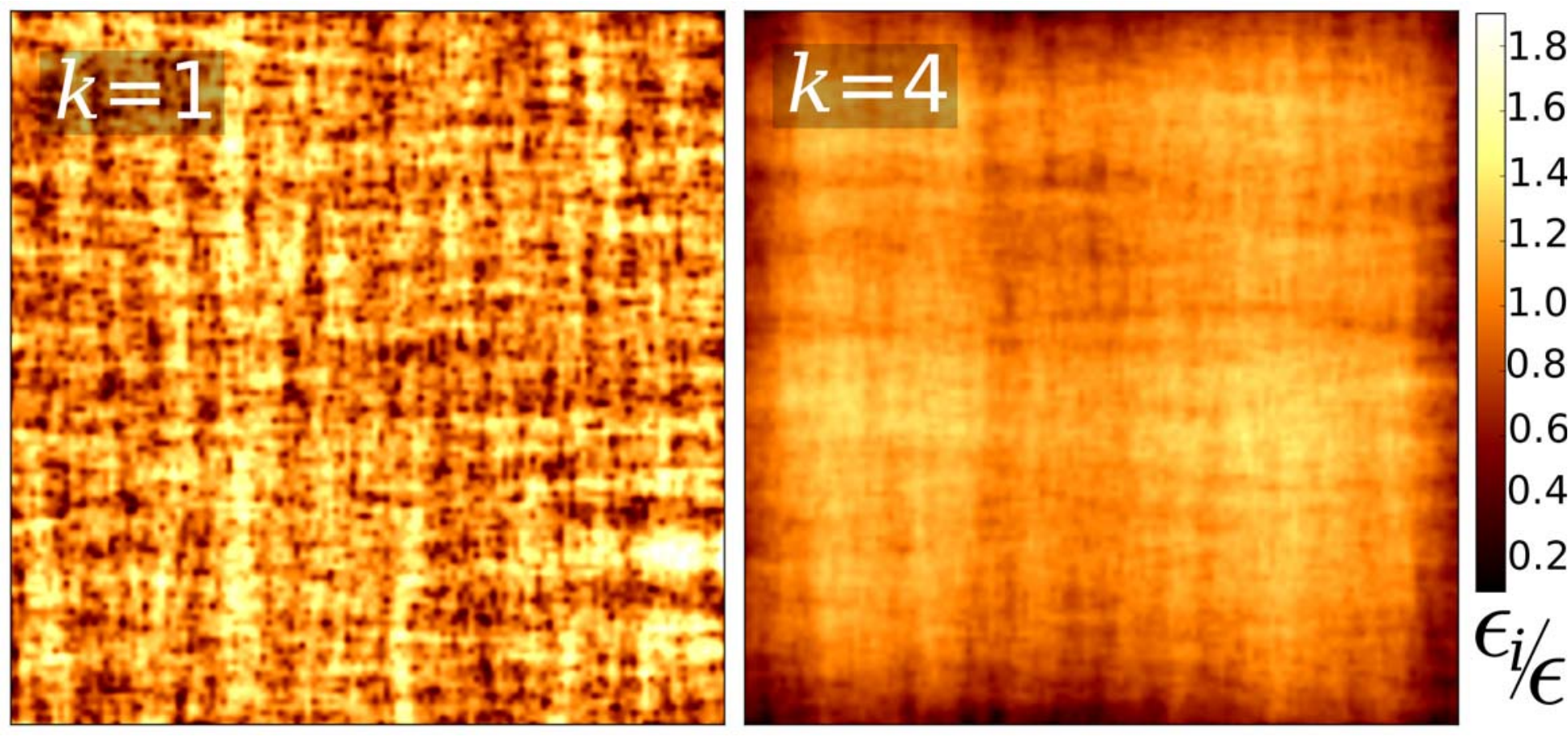}    
    \caption{Strain patterns at $\epsilon=3.0$ for systems with Weibull shape parameter $k=1$ and $k=4$.}
    \label{fig:patterns_diff_k}
\end{figure}

The dependence of $\rho_{\rm II}$ on different system parameters is shown in \figref{corr_coefficient_stationary}. Since spatial correlations are the result of the internal stress field induced by the plastic events, a higher (spatially homogeneous) external stress reduces the effects of such internal stress field and leads to a drop of spatial correlation. Similarly, by increasing temperature we promote the stochastic yielding of elements, leading to reduced spatial correlation. The drop as the system size increases stems from the fact that consecutive events are more likely to occur far away from each other in bigger systems. We observe that, as the disorder increases, the correlation increases. This behavior can be understood by looking at the strain patterns obtained at the same value of strain with different values of $k$. As shown in \figref{patterns_diff_k}, a higher disorder (lower $k$) promotes activity localization in small clusters that are uniformly scattered across the system. In the case of a lower disorder, plastic activity gives rise to comparatively longer-wavelength features, which leads to a lower correlation coefficient.

\subsection{\label{sec:activity_rate} Event rate}

The event rate $\dot n$ is defined as the number of avalanches per unit time. Since avalanches are considered as instantaneous events and are triggered by thermally activated events, $\dot{n}$ matches the rate of thermal activation. 

\subsubsection{Transient regime}
\label{sec:rate_transient}

\begin{figure}
    \includegraphics[width=0.45\textwidth]{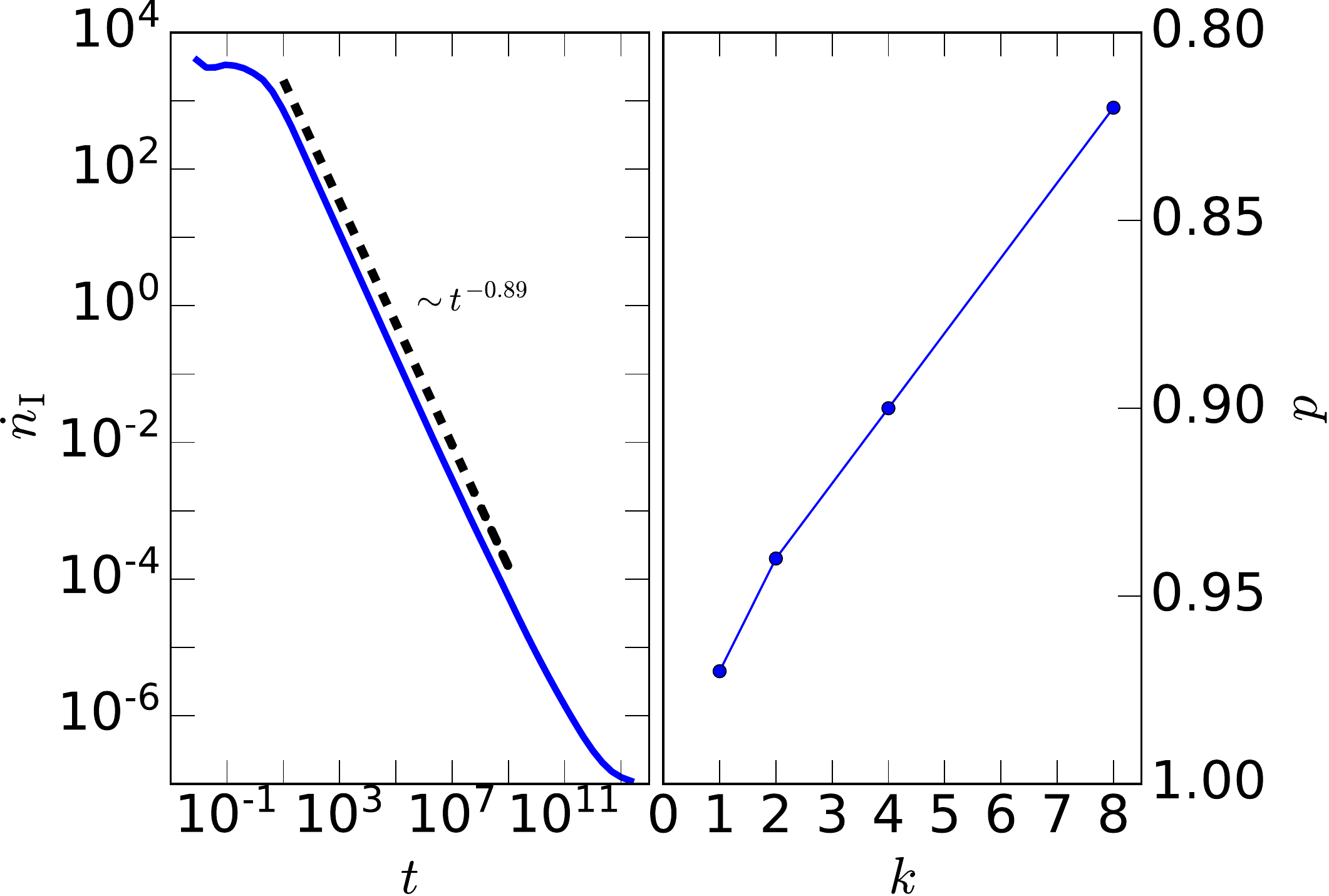}    
    \caption{Left: event rate as a function of time; Right: exponent $p$ of the Omori law as a function of the Weibull shape parameter $k$.}
    \label{fig:omori_law}
\end{figure}

The event rate is found to decrease after the application of the external load as $\dot n \sim t^{-p}$ with $p \approx 0.89$ (see \figref{omori_law}, left). This relation is known as Omori law and describes the decay of activity with time after a large perturbation occurs in the system. It is commonly used in geophysics to model the sequence of aftershocks after a main shock \citep{utsu1995,Schmid2012} and has been found in materials failure tests after a partial rupture of the sample \citep{Baro2013}. In our case, the perturbation corresponds to the sudden application of the external load at $t=0$. The drop of event rate can be explained as the result of the growth of the yield thresholds (\figref{curve_patterns}, bottom) which leads to lower activation rates in \Eqref{rates}. 

We have studied the parameter dependence of the Omori exponent $p$. We find that, within error bars, the value of $p$ does not vary with the value of the applied external stress, the temperature or the system size. However, we find a systematic evolution of the value of $p$ upon variations of the Weibull exponent $k$ (see \figref{omori_law}, right). The exponent $p$ decreases upon increasing the threshold disorder (i.e., lower $k$), and seems to converge towards $p=1$ for $k=0$. Since avalanches have, during our simulations, a constant characteristic size $\langle S \rangle$, the cumulative deformation during the creep process can be obtained by integrating with respect to time the event rate, $\epsilon \sim \int \dot{n} dt$. Consequently, we recover the Andrade law $\epsilon \sim t^{m}$, with an exponent $m = 1-p$, in the case of highly ordered systems of low $k$. On the other hand, as disorder increases and $p \to 1$, we recover logarithmic creep $\epsilon \sim \textrm{log}(t)$. 

\subsubsection{Stationary regime}
\begin{figure}
    \includegraphics[width=0.45\textwidth]{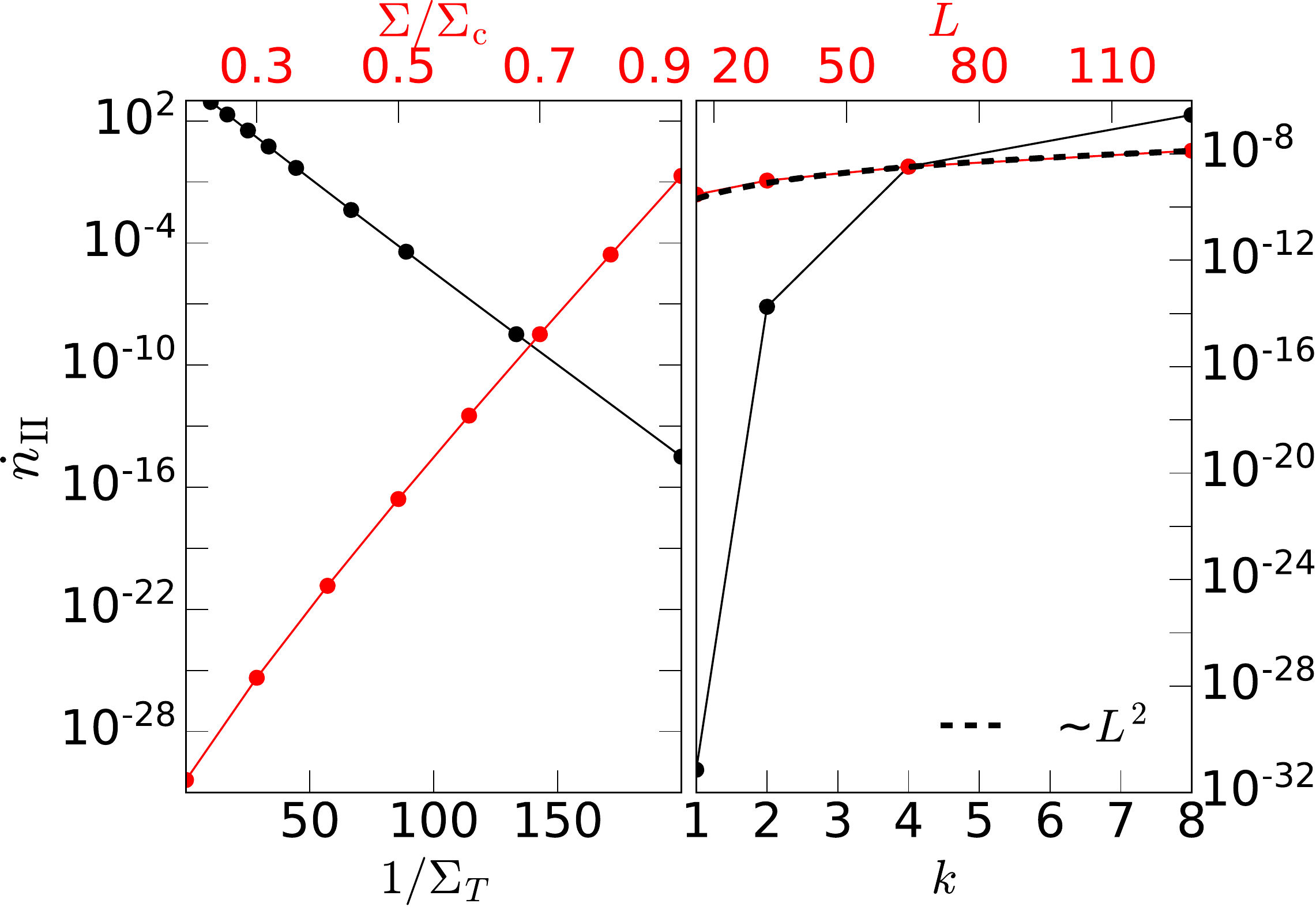}    
    \caption{Activity rate during the stationary creep regime for different system parameters.}
    \label{fig:rate_stationary}
\end{figure}

When the creep process enters the stationary regime, the average threshold becomes constant (\secref{evolution_thresholds}). Since the element activation rates are related to the thresholds through \Eqref{rates} and \Eqref{yield_func}, the activity in the stationary regime $\dot{n}_{\rm II}$ becomes eventually constant. The impact of different simulation parameters on $\dot{n}_{\rm II}$ is shown in \figref{rate_stationary}. The exponential dependence on applied stress and inverse temperature are direct results of the underlying thermal activation process as given by \Eqref{rates}, i.e., our model is in accordance with viscoplastic creep models. The event rate scales with system size as $L^2$, which is the consequence of considering thermal activation a Poisson process in which $L^2$ elements with the same average activation rate (in the stationary regime) attempt to yield simultaneously. Increasing the disorder is found to strongly reduce the event rate in the stationary regime, which is explained by the the higher saturation values of the local strength reached (\figref{av_threshold_stationary}) and again illustrates the paradigm 'more disordered is stronger'.

\subsection{\label{sec:inter-event_time} Inter-event time distributions}

The inter-event time between two thermally activated events is computed by the Kinetic Monte Carlo method. Thus, thermal activation is envisaged as a Poisson process in which the mesoscopic elements attempt to yield independently. However, the distribution of activation barriers is in itself an emergent property as the local activation barriers $\Phi$ depend on the internal stresses and therefore change due to the mutual interactions between events occurring at difffernt times. Therefore, the probability density $P(\Delta t)$ is expected to reflect the different degrees of activity correlation found along the creep curve.

\begin{figure}
	\includegraphics[width=0.45\textwidth]{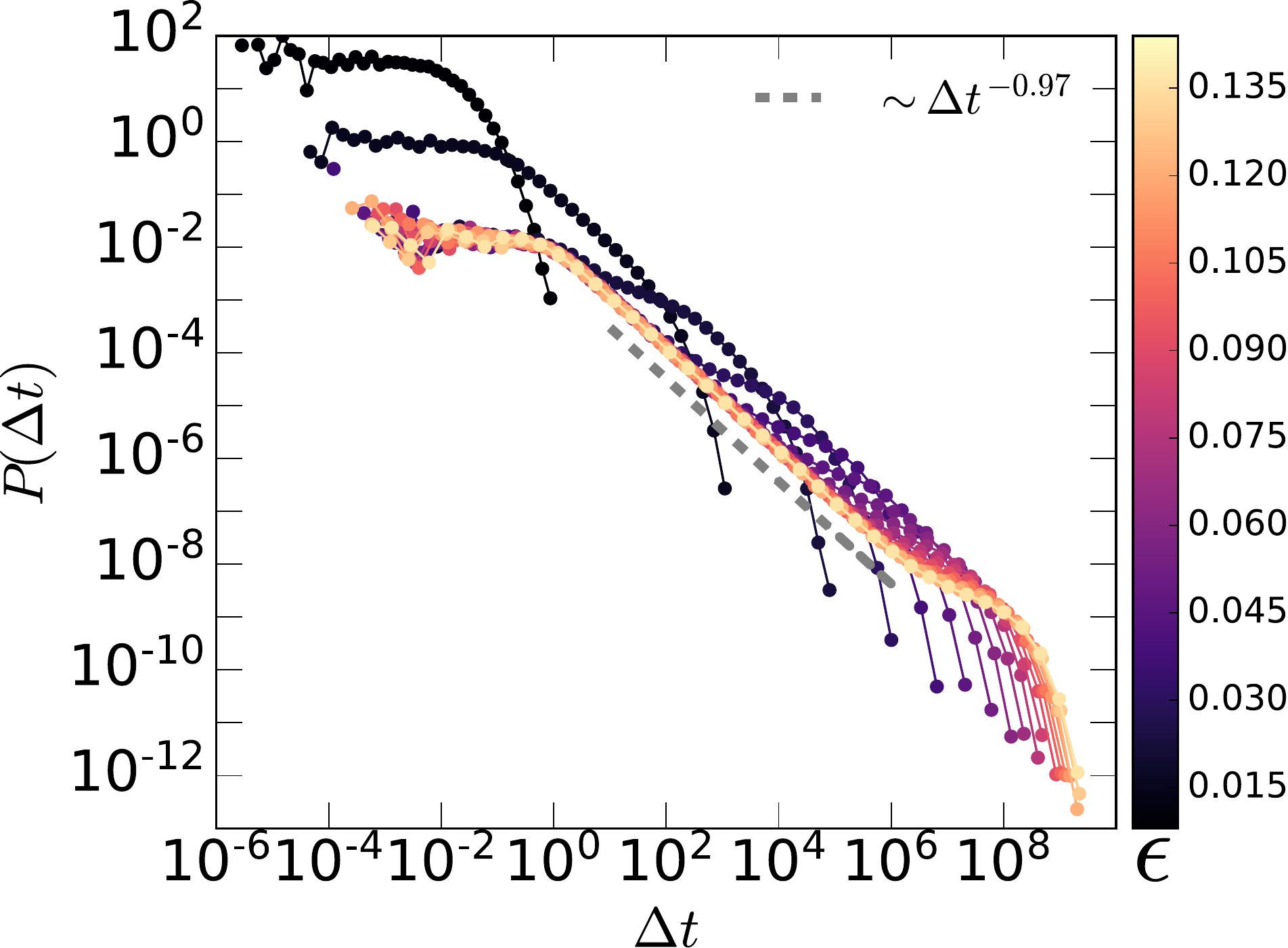}    
	\caption{Inter-event time distribution as a function of strain, for the initial stages of the transient creep regime.}
	\label{fig:waiting_time}
\end{figure}

\figref{waiting_time} shows the evolution of $P(\Delta t)$ during the initial stages of the transient regime. Initially, the probability distribution is of exponential shape, suggesting uncorrelated activity. This is in line with the initial lack of spatial correlation of plastic activity of \secref{event_correlation}. As deformation proceeds, the average threshold grows (\figref{av_threshold_transient_temp}), and the system becomes thus more stable. Consequently, the average time for a thermal fluctuation to occur which can overcome an energy barrier increases, which is reflected in the shift of the distribution cut-off toward bigger values. Still during the transient regime, the probability distribution becomes dominated by a power-law regime with an exponent close to $-1$, incompatible with Poissonian statistics. This happens in parallel with the growth of spatial correlations after the application of the load (\figref{curve_patterns}, bottom). During the stationary creep regime, the distribution of inter-event times becomes stationary and retains its power law charateristics. 

We note that, as shown by \citet{Castellanos2018}, during the third creep stage, i.e. during the approach to failure, the inter-event time distribution undergoes exactly the reverse evolution - it changes from a power law with exponent close to -1 into an exponential distribution, and this evolution is concomitant with a loss of spatial correlation between subsequent events.

\section{\label{sec:discussion}Discussion and Conclusions}

As described in \secref{evolution_thresholds}, the transient creep regime is associated with hardening, reflected in the growth of the average yield threshold, which in turn is the result of the exhaustion of low-threshold elements. The saturation value of the average yield threshold in the stationary creep regime is found to depend on the system parameters (\figref{av_threshold_stationary}). At low temperatures, low threshold elements are typically chosen for yielding. The renewed threshold is thus likely to be significantly higher than the previous one, which enhances statistical hardening. On the other hand, a higher temperature allows for more stable elements to be chosen for yielding, reducing thus the effects of statistical hardening. The stationary threshold is determined, at a certain temperature, by the dynamical equilibrium achived between both processes, namely the seletion of low-threshold elements which become stronger after yielding and the selection of strong elements which become softer after yielding. Increasing the external stress has an effect similar to that of  increasing temperature. This can be understood by noticing that the stress lowers the activation barriers in a global fashion. Consequently, at a higher external stress, strong elements are chosen for yielding more often than they would at low external stress. This has the effect of shifting the dynamical equilibrium of thresholds toward lower values. By increasing the threshold disorder (i.e., reducing $k$), the stationary average yield threshold is increased, which is a direct consequence of fattening the high-strength tail of the threshold probability distribution. 

We find that the decrease of activity during the transient creep regime follows an Omori-like law with an exponent $p$ which is generally close to $1$, as is experimentally well-known for a wide range of materials, and recently established in the specific case of metallic glasses \cite{McFaul2018}. Specifically, we observe that $p$ does not depend on the external stress, temperature or system size but varies systematically with the degree of disorder. The exponent seems to converge towards $p=1$ at high disorder, while the value decreases with the disorder. As pointed out in \secref{rate_transient}, the variation of $p$ with microstructural disorder might explain the variability observed in the exponent of the Andrade law of transient creep or the existence of logarithmic creep. Moreover, the exclusive dependence of the exponent on the disorder of the yield thresholds might allow for the quantitative estimation of the disorder at the mesoscopic scale. The disorder at the mesoscale is a crucial parameter in the modeling of the plastic activity of disordered materials \cite{Tuszes2017,BudrikisNatCom,wyart2018,Ozawa2018}, which is on the other hand difficult to establish experimentally. It is therefore of interest establishing links between the macroscopic observable response and a quantitative measure of structural heterogeneity.
 
The evolution of the inter-event time distribution as deformation proceeds (\figref{waiting_time}) is characterized by a transition from an initially exponential shape to a power-law-like distribution with an exponent of approximately $-1$. This value is in agreement with experimental findings on the creep deformation of bulk metallic glasses \cite{Krisponeit2014_NatCom,McFaul2018}. The transition from uncorrelated to correlated activity suggested by the distributions of inter-event times is supported by the correlation coefficient of spatial activity (\secref{event_correlation}). Interestingly, \citet{Krisponeit2014_NatCom} found that the distribution of inter-event times evolves with time, from a distribution initially dominated by a cut-off towards a power-law-like distribution with an exponent of approximately $-0.8 \pm 0.1$, which agrees well with our observations. 

In conclusion, we have presented a mesoscale elastoplastic model which can reproduce the time evolution of the intermittent plastic activity of disordered materials under creep conditions. We find that the structural evolution as a function of time is concomitant to a variation of the spatial correlation of events, to the emergence of strain localization patterns, and to the variation of the event rate and the inter-event times distribution. In summary, our approach allows us to study the creep dynamics from a holistic perspective, establishing contact between diverse approaches which are frequently considered separately, the macroscopic smooth flow response, the microscopic stochastic intermittent response, and the evolution of the structural properties as expressed by the distribution of local activation thresholds.

\bibliography{paper}

\begin{thebibliography}{46}%
\makeatletter
\providecommand \@ifxundefined [1]{%
 \@ifx{#1\undefined}
}%
\providecommand \@ifnum [1]{%
 \ifnum #1\expandafter \@firstoftwo
 \else \expandafter \@secondoftwo
 \fi
}%
\providecommand \@ifx [1]{%
 \ifx #1\expandafter \@firstoftwo
 \else \expandafter \@secondoftwo
 \fi
}%
\providecommand \natexlab [1]{#1}%
\providecommand \enquote  [1]{``#1''}%
\providecommand \bibnamefont  [1]{#1}%
\providecommand \bibfnamefont [1]{#1}%
\providecommand \citenamefont [1]{#1}%
\providecommand \href@noop [0]{\@secondoftwo}%
\providecommand \href [0]{\begingroup \@sanitize@url \@href}%
\providecommand \@href[1]{\@@startlink{#1}\@@href}%
\providecommand \@@href[1]{\endgroup#1\@@endlink}%
\providecommand \@sanitize@url [0]{\catcode `\\12\catcode `\$12\catcode
  `\&12\catcode `\#12\catcode `\^12\catcode `\_12\catcode `\%12\relax}%
\providecommand \@@startlink[1]{}%
\providecommand \@@endlink[0]{}%
\providecommand \url  [0]{\begingroup\@sanitize@url \@url }%
\providecommand \@url [1]{\endgroup\@href {#1}{\urlprefix }}%
\providecommand \urlprefix  [0]{URL }%
\providecommand \Eprint [0]{\href }%
\providecommand \doibase [0]{http://dx.doi.org/}%
\providecommand \selectlanguage [0]{\@gobble}%
\providecommand \bibinfo  [0]{\@secondoftwo}%
\providecommand \bibfield  [0]{\@secondoftwo}%
\providecommand \translation [1]{[#1]}%
\providecommand \BibitemOpen [0]{}%
\providecommand \bibitemStop [0]{}%
\providecommand \bibitemNoStop [0]{.\EOS\space}%
\providecommand \EOS [0]{\spacefactor3000\relax}%
\providecommand \BibitemShut  [1]{\csname bibitem#1\endcsname}%
\let\auto@bib@innerbib\@empty
\bibitem [{\citenamefont {M.Sc.}(1905)}]{Phillips1905}%
  \BibitemOpen
  \bibfield  {author} {\bibinfo {author} {\bibfnamefont {P.~P.}\ \bibnamefont
  {M.Sc.}},\ }\href {\doibase 10.1080/14786440509463303} {\bibfield  {journal}
  {\bibinfo  {journal} {The London, Edinburgh, and Dublin Philosophical
  Magazine and Journal of Science}\ }\textbf {\bibinfo {volume} {9}},\ \bibinfo
  {pages} {513} (\bibinfo {year} {1905})},\ \Eprint
  {http://arxiv.org/abs/https://doi.org/10.1080/14786440509463303}
  {https://doi.org/10.1080/14786440509463303} \BibitemShut {NoStop}%
\bibitem [{\citenamefont {Andrade}\ and\ \citenamefont
  {Trouton}(1910)}]{Andrade1910}%
  \BibitemOpen
  \bibfield  {author} {\bibinfo {author} {\bibfnamefont {E.~N. D.~C.}\
  \bibnamefont {Andrade}}\ and\ \bibinfo {author} {\bibfnamefont {F.~T.}\
  \bibnamefont {Trouton}},\ }\href {\doibase 10.1098/rspa.1910.0050} {\bibfield
   {journal} {\bibinfo  {journal} {Proceedings of the Royal Society of London.
  Series A, Containing Papers of a Mathematical and Physical Character}\
  }\textbf {\bibinfo {volume} {84}},\ \bibinfo {pages} {1} (\bibinfo {year}
  {1910})},\ \Eprint
  {http://arxiv.org/abs/https://royalsocietypublishing.org/doi/pdf/10.1098/rspa.1910.0050}
  {https://royalsocietypublishing.org/doi/pdf/10.1098/rspa.1910.0050}
  \BibitemShut {NoStop}%
\bibitem [{\citenamefont {Lee}\ \emph {et~al.}(1991)\citenamefont {Lee},
  \citenamefont {Liu},\ and\ \citenamefont {Chen}}]{Lee1991}%
  \BibitemOpen
  \bibfield  {author} {\bibinfo {author} {\bibfnamefont {H.~M.}\ \bibnamefont
  {Lee}}, \bibinfo {author} {\bibfnamefont {X.~L.}\ \bibnamefont {Liu}}, \ and\
  \bibinfo {author} {\bibfnamefont {W.~F.}\ \bibnamefont {Chen}},\ }\href
  {\doibase 10.1061/(ASCE)0733-9445(1991)117:10(3135)} {\bibfield  {journal}
  {\bibinfo  {journal} {Journal of Structural Engineering}\ }\textbf {\bibinfo
  {volume} {117}},\ \bibinfo {pages} {3135} (\bibinfo {year}
  {1991})}\BibitemShut {NoStop}%
\bibitem [{\citenamefont {Sleep}\ and\ \citenamefont
  {Blanpied}(1992)}]{Sleep1992}%
  \BibitemOpen
  \bibfield  {author} {\bibinfo {author} {\bibfnamefont {N.~H.}\ \bibnamefont
  {Sleep}}\ and\ \bibinfo {author} {\bibfnamefont {M.~L.}\ \bibnamefont
  {Blanpied}},\ }\href {\doibase 10.1038/359687a0} {\bibfield  {journal}
  {\bibinfo  {journal} {Nature}\ }\textbf {\bibinfo {volume} {359}},\ \bibinfo
  {pages} {687} (\bibinfo {year} {1992})}\BibitemShut {NoStop}%
\bibitem [{\citenamefont {Louchet}\ and\ \citenamefont
  {Duval}(2009)}]{Louchet2009}%
  \BibitemOpen
  \bibfield  {author} {\bibinfo {author} {\bibfnamefont {F.}~\bibnamefont
  {Louchet}}\ and\ \bibinfo {author} {\bibfnamefont {P.}~\bibnamefont
  {Duval}},\ }\href {\doibase 10.3139/146.110189} {\bibfield  {journal}
  {\bibinfo  {journal} {International Journal of Materials Research}\ }\textbf
  {\bibinfo {volume} {100}},\ \bibinfo {pages} {1433} (\bibinfo {year}
  {2009})}\BibitemShut {NoStop}%
\bibitem [{\citenamefont {Cottrell}(1997)}]{Cottrell1997}%
  \BibitemOpen
  \bibfield  {author} {\bibinfo {author} {\bibfnamefont {A.~H.}\ \bibnamefont
  {Cottrell}},\ }\href {\doibase 10.1080/095008397179552} {\bibfield  {journal}
  {\bibinfo  {journal} {Philosophical Magazine Letters}\ }\textbf {\bibinfo
  {volume} {75}},\ \bibinfo {pages} {301} (\bibinfo {year} {1997})},\ \Eprint
  {http://arxiv.org/abs/https://doi.org/10.1080/095008397179552}
  {https://doi.org/10.1080/095008397179552} \BibitemShut {NoStop}%
\bibitem [{\citenamefont {Main}(2000)}]{Main2000_GJI}%
  \BibitemOpen
  \bibfield  {author} {\bibinfo {author} {\bibfnamefont {I.~G.}\ \bibnamefont
  {Main}},\ }\href {\doibase 10.1046/j.1365-246x.2000.00136.x} {\bibfield
  {journal} {\bibinfo  {journal} {Geophysical Journal International}\ }\textbf
  {\bibinfo {volume} {142}},\ \bibinfo {pages} {151} (\bibinfo {year}
  {2000})},\ \Eprint
  {http://arxiv.org/abs/http://gji.oxfordjournals.org/content/142/1/151.full.pdf+html}
  {http://gji.oxfordjournals.org/content/142/1/151.full.pdf+html} \BibitemShut
  {NoStop}%
\bibitem [{\citenamefont {Zaiser}(2013)}]{Zaiser2008}%
  \BibitemOpen
  \bibfield  {author} {\bibinfo {author} {\bibfnamefont {M.}~\bibnamefont
  {Zaiser}},\ }\href@noop {} {\bibfield  {journal} {\bibinfo  {journal} {J Mech
  Behav Mater}\ }\textbf {\bibinfo {volume} {22}},\ \bibinfo {pages} {3}
  (\bibinfo {year} {2013})}\BibitemShut {NoStop}%
\bibitem [{\citenamefont {Schneider}\ \emph {et~al.}(2009)\citenamefont
  {Schneider}, \citenamefont {Clark}, \citenamefont {Frick}, \citenamefont
  {Gruber},\ and\ \citenamefont {Arzt}}]{Schneider2009}%
  \BibitemOpen
  \bibfield  {author} {\bibinfo {author} {\bibfnamefont {A.}~\bibnamefont
  {Schneider}}, \bibinfo {author} {\bibfnamefont {B.}~\bibnamefont {Clark}},
  \bibinfo {author} {\bibfnamefont {C.}~\bibnamefont {Frick}}, \bibinfo
  {author} {\bibfnamefont {P.}~\bibnamefont {Gruber}}, \ and\ \bibinfo {author}
  {\bibfnamefont {E.}~\bibnamefont {Arzt}},\ }\href {\doibase
  https://doi.org/10.1016/j.msea.2009.01.011} {\bibfield  {journal} {\bibinfo
  {journal} {Materials Science and Engineering: A}\ }\textbf {\bibinfo {volume}
  {508}},\ \bibinfo {pages} {241 } (\bibinfo {year} {2009})}\BibitemShut
  {NoStop}%
\bibitem [{\citenamefont {Omori}(1894)}]{Omori1894}%
  \BibitemOpen
  \bibfield  {author} {\bibinfo {author} {\bibfnamefont {F.}~\bibnamefont
  {Omori}},\ }\href@noop {} {\bibfield  {journal} {\bibinfo  {journal} {J.
  Coll. Sci. Imp. Univ. Tokyo}\ }\textbf {\bibinfo {volume} {7}},\ \bibinfo
  {pages} {111} (\bibinfo {year} {1894})}\BibitemShut {NoStop}%
\bibitem [{\citenamefont {Utsu}\ \emph {et~al.}(1995)\citenamefont {Utsu},
  \citenamefont {Ogata}, \citenamefont {S},\ and\ \citenamefont
  {Matsu'ura}}]{utsu1995}%
  \BibitemOpen
  \bibfield  {author} {\bibinfo {author} {\bibfnamefont {T.}~\bibnamefont
  {Utsu}}, \bibinfo {author} {\bibfnamefont {Y.}~\bibnamefont {Ogata}},
  \bibinfo {author} {\bibfnamefont {R.}~\bibnamefont {S}}, \ and\ \bibinfo
  {author} {\bibnamefont {Matsu'ura}},\ }\href {\doibase 10.4294/jpe1952.43.1}
  {\bibfield  {journal} {\bibinfo  {journal} {Journal of Physics of the Earth}\
  }\textbf {\bibinfo {volume} {43}},\ \bibinfo {pages} {1} (\bibinfo {year}
  {1995})}\BibitemShut {NoStop}%
\bibitem [{\citenamefont {Lennartz-Sassinek}\ \emph {et~al.}(2014)\citenamefont
  {Lennartz-Sassinek}, \citenamefont {Main}, \citenamefont {Zaiser},\ and\
  \citenamefont {Graham}}]{Lennartz-Sassinek2014}%
  \BibitemOpen
  \bibfield  {author} {\bibinfo {author} {\bibfnamefont {S.}~\bibnamefont
  {Lennartz-Sassinek}}, \bibinfo {author} {\bibfnamefont {I.~G.}\ \bibnamefont
  {Main}}, \bibinfo {author} {\bibfnamefont {M.}~\bibnamefont {Zaiser}}, \ and\
  \bibinfo {author} {\bibfnamefont {C.~C.}\ \bibnamefont {Graham}},\ }\href
  {\doibase 10.1103/PhysRevE.90.052401} {\bibfield  {journal} {\bibinfo
  {journal} {Phys. Rev. E}\ }\textbf {\bibinfo {volume} {90}},\ \bibinfo
  {pages} {052401} (\bibinfo {year} {2014})}\BibitemShut {NoStop}%
\bibitem [{\citenamefont {Bar\'o}\ \emph {et~al.}(2013)\citenamefont {Bar\'o},
  \citenamefont {Corral}, \citenamefont {Illa}, \citenamefont {Planes},
  \citenamefont {Salje}, \citenamefont {Schranz}, \citenamefont {Soto-Parra},\
  and\ \citenamefont {Vives}}]{Baro2013}%
  \BibitemOpen
  \bibfield  {author} {\bibinfo {author} {\bibfnamefont {J.}~\bibnamefont
  {Bar\'o}}, \bibinfo {author} {\bibfnamefont {A.}~\bibnamefont {Corral}},
  \bibinfo {author} {\bibfnamefont {X.}~\bibnamefont {Illa}}, \bibinfo {author}
  {\bibfnamefont {A.}~\bibnamefont {Planes}}, \bibinfo {author} {\bibfnamefont
  {E.~K.~H.}\ \bibnamefont {Salje}}, \bibinfo {author} {\bibfnamefont
  {W.}~\bibnamefont {Schranz}}, \bibinfo {author} {\bibfnamefont {D.~E.}\
  \bibnamefont {Soto-Parra}}, \ and\ \bibinfo {author} {\bibfnamefont
  {E.}~\bibnamefont {Vives}},\ }\href {\doibase 10.1103/PhysRevLett.110.088702}
  {\bibfield  {journal} {\bibinfo  {journal} {Phys. Rev. Lett.}\ }\textbf
  {\bibinfo {volume} {110}},\ \bibinfo {pages} {088702} (\bibinfo {year}
  {2013})}\BibitemShut {NoStop}%
\bibitem [{\citenamefont {Leocmach}\ \emph {et~al.}(2014)\citenamefont
  {Leocmach}, \citenamefont {Perge}, \citenamefont {Divoux},\ and\
  \citenamefont {Manneville}}]{Leocmach2014}%
  \BibitemOpen
  \bibfield  {author} {\bibinfo {author} {\bibfnamefont {M.}~\bibnamefont
  {Leocmach}}, \bibinfo {author} {\bibfnamefont {C.}~\bibnamefont {Perge}},
  \bibinfo {author} {\bibfnamefont {T.}~\bibnamefont {Divoux}}, \ and\ \bibinfo
  {author} {\bibfnamefont {S.}~\bibnamefont {Manneville}},\ }\href {\doibase
  10.1103/PhysRevLett.113.038303} {\bibfield  {journal} {\bibinfo  {journal}
  {Phys. Rev. Lett.}\ }\textbf {\bibinfo {volume} {113}},\ \bibinfo {pages}
  {038303} (\bibinfo {year} {2014})}\BibitemShut {NoStop}%
\bibitem [{\citenamefont {McFaul}\ \emph {et~al.}(2018)\citenamefont {McFaul},
  \citenamefont {Wright}, \citenamefont {Gu}, \citenamefont {Uhl},\ and\
  \citenamefont {Dahmen}}]{McFaul2018}%
  \BibitemOpen
  \bibfield  {author} {\bibinfo {author} {\bibfnamefont {L.~W.}\ \bibnamefont
  {McFaul}}, \bibinfo {author} {\bibfnamefont {W.~J.}\ \bibnamefont {Wright}},
  \bibinfo {author} {\bibfnamefont {X.}~\bibnamefont {Gu}}, \bibinfo {author}
  {\bibfnamefont {J.~T.}\ \bibnamefont {Uhl}}, \ and\ \bibinfo {author}
  {\bibfnamefont {K.~A.}\ \bibnamefont {Dahmen}},\ }\href {\doibase
  10.1103/PhysRevE.97.063005} {\bibfield  {journal} {\bibinfo  {journal} {Phys.
  Rev. E}\ }\textbf {\bibinfo {volume} {97}},\ \bibinfo {pages} {063005}
  (\bibinfo {year} {2018})}\BibitemShut {NoStop}%
\bibitem [{\citenamefont {Cottrell}(2004)}]{Cotrell2004}%
  \BibitemOpen
  \bibfield  {author} {\bibinfo {author} {\bibfnamefont {A.~H.}\ \bibnamefont
  {Cottrell}},\ }\href {\doibase 10.1080/09500830500036146} {\bibfield
  {journal} {\bibinfo  {journal} {Philosophical Magazine Letters}\ }\textbf
  {\bibinfo {volume} {84}},\ \bibinfo {pages} {685} (\bibinfo {year} {2004})},\
  \Eprint {http://arxiv.org/abs/https://doi.org/10.1080/09500830500036146}
  {https://doi.org/10.1080/09500830500036146} \BibitemShut {NoStop}%
\bibitem [{\citenamefont {Miguel}\ \emph {et~al.}(2002)\citenamefont {Miguel},
  \citenamefont {Vespignani}, \citenamefont {Zaiser},\ and\ \citenamefont
  {Zapperi}}]{Miguel2002}%
  \BibitemOpen
  \bibfield  {author} {\bibinfo {author} {\bibfnamefont {M.-C.}\ \bibnamefont
  {Miguel}}, \bibinfo {author} {\bibfnamefont {A.}~\bibnamefont {Vespignani}},
  \bibinfo {author} {\bibfnamefont {M.}~\bibnamefont {Zaiser}}, \ and\ \bibinfo
  {author} {\bibfnamefont {S.}~\bibnamefont {Zapperi}},\ }\href {\doibase
  10.1103/PhysRevLett.89.165501} {\bibfield  {journal} {\bibinfo  {journal}
  {Phys. Rev. Lett.}\ }\textbf {\bibinfo {volume} {89}},\ \bibinfo {pages}
  {165501} (\bibinfo {year} {2002})}\BibitemShut {NoStop}%
\bibitem [{\citenamefont {Zaiser}\ and\ \citenamefont
  {Hähne}(1999)}]{Zaiser1999}%
  \BibitemOpen
  \bibfield  {author} {\bibinfo {author} {\bibfnamefont {M.}~\bibnamefont
  {Zaiser}}\ and\ \bibinfo {author} {\bibfnamefont {P.}~\bibnamefont
  {Hähne}},\ }\href@noop {} {\bibfield  {journal} {\bibinfo  {journal} {Mater.
  Sci. Engng. A}\ }\textbf {\bibinfo {volume} {270}},\ \bibinfo {pages} {2999}
  (\bibinfo {year} {1999})}\BibitemShut {NoStop}%
\bibitem [{\citenamefont {Zaiser}\ and\ \citenamefont
  {Aifantis}(2006)}]{Zaiser2006}%
  \BibitemOpen
  \bibfield  {author} {\bibinfo {author} {\bibfnamefont {M.}~\bibnamefont
  {Zaiser}}\ and\ \bibinfo {author} {\bibfnamefont {E.}~\bibnamefont
  {Aifantis}},\ }\href@noop {} {\bibfield  {journal} {\bibinfo  {journal} {Int.
  J. Plasticity}\ }\textbf {\bibinfo {volume} {22}},\ \bibinfo {pages} {1432}
  (\bibinfo {year} {2006})}\BibitemShut {NoStop}%
\bibitem [{\citenamefont {Argon}(1979)}]{Argon1979}%
  \BibitemOpen
  \bibfield  {author} {\bibinfo {author} {\bibfnamefont {A.}~\bibnamefont
  {Argon}},\ }\href {\doibase http://dx.doi.org/10.1016/0001-6160(79)90055-5}
  {\bibfield  {journal} {\bibinfo  {journal} {Acta Metallurgica}\ }\textbf
  {\bibinfo {volume} {27}},\ \bibinfo {pages} {47 } (\bibinfo {year}
  {1979})}\BibitemShut {NoStop}%
\bibitem [{\citenamefont {Schuh}\ \emph {et~al.}(2007)\citenamefont {Schuh},
  \citenamefont {Hufnagel},\ and\ \citenamefont {Ramamurty}}]{Schuh2007}%
  \BibitemOpen
  \bibfield  {author} {\bibinfo {author} {\bibfnamefont {C.~A.}\ \bibnamefont
  {Schuh}}, \bibinfo {author} {\bibfnamefont {T.~C.}\ \bibnamefont {Hufnagel}},
  \ and\ \bibinfo {author} {\bibfnamefont {U.}~\bibnamefont {Ramamurty}},\
  }\href {\doibase http://dx.doi.org/10.1016/j.actamat.2007.01.052} {\bibfield
  {journal} {\bibinfo  {journal} {Acta Materialia}\ }\textbf {\bibinfo {volume}
  {55}},\ \bibinfo {pages} {4067 } (\bibinfo {year} {2007})}\BibitemShut
  {NoStop}%
\bibitem [{\citenamefont {Krisponeit}\ \emph {et~al.}(2014)\citenamefont
  {Krisponeit}, \citenamefont {Pitikaris}, \citenamefont {Avila}, \citenamefont
  {K{\"u}chemann}, \citenamefont {Kr{\"u}ger},\ and\ \citenamefont
  {Samwer}}]{Krisponeit2014_NatCom}%
  \BibitemOpen
  \bibfield  {author} {\bibinfo {author} {\bibfnamefont {J.-O.}\ \bibnamefont
  {Krisponeit}}, \bibinfo {author} {\bibfnamefont {S.}~\bibnamefont
  {Pitikaris}}, \bibinfo {author} {\bibfnamefont {K.~E.}\ \bibnamefont
  {Avila}}, \bibinfo {author} {\bibfnamefont {S.}~\bibnamefont
  {K{\"u}chemann}}, \bibinfo {author} {\bibfnamefont {A.}~\bibnamefont
  {Kr{\"u}ger}}, \ and\ \bibinfo {author} {\bibfnamefont {K.}~\bibnamefont
  {Samwer}},\ }\href {\doibase 10.1038/ncomms4616} {\ \textbf {\bibinfo
  {volume} {5}},\ \bibinfo {pages} {3616 EP } (\bibinfo {year} {2014})},\
  \bibinfo {note} {article}\BibitemShut {NoStop}%
\bibitem [{\citenamefont {Heap}\ \emph {et~al.}(2011)\citenamefont {Heap},
  \citenamefont {Baud}, \citenamefont {Meredith}, \citenamefont {Vinciguerra},
  \citenamefont {Bell},\ and\ \citenamefont {Main}}]{Heap2011_EPSL}%
  \BibitemOpen
  \bibfield  {author} {\bibinfo {author} {\bibfnamefont {M.}~\bibnamefont
  {Heap}}, \bibinfo {author} {\bibfnamefont {P.}~\bibnamefont {Baud}}, \bibinfo
  {author} {\bibfnamefont {P.}~\bibnamefont {Meredith}}, \bibinfo {author}
  {\bibfnamefont {S.}~\bibnamefont {Vinciguerra}}, \bibinfo {author}
  {\bibfnamefont {A.}~\bibnamefont {Bell}}, \ and\ \bibinfo {author}
  {\bibfnamefont {I.}~\bibnamefont {Main}},\ }\href {\doibase
  http://dx.doi.org/10.1016/j.epsl.2011.04.035} {\bibfield  {journal} {\bibinfo
   {journal} {Earth Planet. Sci. Lett.}\ }\textbf {\bibinfo {volume} {307}},\
  \bibinfo {pages} {71 } (\bibinfo {year} {2011})}\BibitemShut {NoStop}%
\bibitem [{\citenamefont {Rosti}\ \emph {et~al.}(2010)\citenamefont {Rosti},
  \citenamefont {Koivisto}, \citenamefont {Laurson},\ and\ \citenamefont
  {Alava}}]{Rosti2010_PRL}%
  \BibitemOpen
  \bibfield  {author} {\bibinfo {author} {\bibfnamefont {J.}~\bibnamefont
  {Rosti}}, \bibinfo {author} {\bibfnamefont {J.}~\bibnamefont {Koivisto}},
  \bibinfo {author} {\bibfnamefont {L.}~\bibnamefont {Laurson}}, \ and\
  \bibinfo {author} {\bibfnamefont {M.~J.}\ \bibnamefont {Alava}},\ }\href
  {\doibase 10.1103/PhysRevLett.105.100601} {\bibfield  {journal} {\bibinfo
  {journal} {Phys. Rev. Lett.}\ }\textbf {\bibinfo {volume} {105}},\ \bibinfo
  {pages} {100601} (\bibinfo {year} {2010})}\BibitemShut {NoStop}%
\bibitem [{\citenamefont {Nguyen}\ \emph {et~al.}(2011)\citenamefont {Nguyen},
  \citenamefont {Darnige}, \citenamefont {Bruand},\ and\ \citenamefont
  {Clement}}]{Nguyen2011}%
  \BibitemOpen
  \bibfield  {author} {\bibinfo {author} {\bibfnamefont {V.~B.}\ \bibnamefont
  {Nguyen}}, \bibinfo {author} {\bibfnamefont {T.}~\bibnamefont {Darnige}},
  \bibinfo {author} {\bibfnamefont {A.}~\bibnamefont {Bruand}}, \ and\ \bibinfo
  {author} {\bibfnamefont {E.}~\bibnamefont {Clement}},\ }\href {\doibase
  10.1103/PhysRevLett.107.138303} {\bibfield  {journal} {\bibinfo  {journal}
  {Phys. Rev. Lett.}\ }\textbf {\bibinfo {volume} {107}},\ \bibinfo {pages}
  {138303} (\bibinfo {year} {2011})}\BibitemShut {NoStop}%
\bibitem [{\citenamefont {Nicolas}\ \emph {et~al.}(2018)\citenamefont
  {Nicolas}, \citenamefont {Ferrero}, \citenamefont {Martens},\ and\
  \citenamefont {Barrat}}]{Nicolas2018}%
  \BibitemOpen
  \bibfield  {author} {\bibinfo {author} {\bibfnamefont {A.}~\bibnamefont
  {Nicolas}}, \bibinfo {author} {\bibfnamefont {E.~E.}\ \bibnamefont
  {Ferrero}}, \bibinfo {author} {\bibfnamefont {K.}~\bibnamefont {Martens}}, \
  and\ \bibinfo {author} {\bibfnamefont {J.-L.}\ \bibnamefont {Barrat}},\
  }\href {\doibase 10.1103/RevModPhys.90.045006} {\bibfield  {journal}
  {\bibinfo  {journal} {Rev. Mod. Phys.}\ }\textbf {\bibinfo {volume} {90}},\
  \bibinfo {pages} {045006} (\bibinfo {year} {2018})}\BibitemShut {NoStop}%
\bibitem [{\citenamefont {Barbot}\ \emph {et~al.}(2018)\citenamefont {Barbot},
  \citenamefont {Lerbinger}, \citenamefont {Hernandez-Garcia}, \citenamefont
  {Garc\'{\i}a-Garc\'{\i}a}, \citenamefont {Falk}, \citenamefont
  {Vandembroucq},\ and\ \citenamefont {Patinet}}]{Barbot2018}%
  \BibitemOpen
  \bibfield  {author} {\bibinfo {author} {\bibfnamefont {A.}~\bibnamefont
  {Barbot}}, \bibinfo {author} {\bibfnamefont {M.}~\bibnamefont {Lerbinger}},
  \bibinfo {author} {\bibfnamefont {A.}~\bibnamefont {Hernandez-Garcia}},
  \bibinfo {author} {\bibfnamefont {R.}~\bibnamefont
  {Garc\'{\i}a-Garc\'{\i}a}}, \bibinfo {author} {\bibfnamefont {M.~L.}\
  \bibnamefont {Falk}}, \bibinfo {author} {\bibfnamefont {D.}~\bibnamefont
  {Vandembroucq}}, \ and\ \bibinfo {author} {\bibfnamefont {S.}~\bibnamefont
  {Patinet}},\ }\href {\doibase 10.1103/PhysRevE.97.033001} {\bibfield
  {journal} {\bibinfo  {journal} {Phys. Rev. E}\ }\textbf {\bibinfo {volume}
  {97}},\ \bibinfo {pages} {033001} (\bibinfo {year} {2018})}\BibitemShut
  {NoStop}%
\bibitem [{\citenamefont {Patinet}\ \emph {et~al.}(2016)\citenamefont
  {Patinet}, \citenamefont {Vandembroucq},\ and\ \citenamefont
  {Falk}}]{Patinet2016}%
  \BibitemOpen
  \bibfield  {author} {\bibinfo {author} {\bibfnamefont {S.}~\bibnamefont
  {Patinet}}, \bibinfo {author} {\bibfnamefont {D.}~\bibnamefont
  {Vandembroucq}}, \ and\ \bibinfo {author} {\bibfnamefont {M.~L.}\
  \bibnamefont {Falk}},\ }\href {\doibase 10.1103/PhysRevLett.117.045501}
  {\bibfield  {journal} {\bibinfo  {journal} {Phys. Rev. Lett.}\ }\textbf
  {\bibinfo {volume} {117}},\ \bibinfo {pages} {045501} (\bibinfo {year}
  {2016})}\BibitemShut {NoStop}%
\bibitem [{\citenamefont {Budrikis}\ \emph {et~al.}(2017)\citenamefont
  {Budrikis}, \citenamefont {Castellanos}, \citenamefont {Sandfeld},
  \citenamefont {Zaiser},\ and\ \citenamefont {Zapperi}}]{BudrikisNatCom}%
  \BibitemOpen
  \bibfield  {author} {\bibinfo {author} {\bibfnamefont {Z.}~\bibnamefont
  {Budrikis}}, \bibinfo {author} {\bibfnamefont {D.~F.}\ \bibnamefont
  {Castellanos}}, \bibinfo {author} {\bibfnamefont {S.}~\bibnamefont
  {Sandfeld}}, \bibinfo {author} {\bibfnamefont {M.}~\bibnamefont {Zaiser}}, \
  and\ \bibinfo {author} {\bibfnamefont {S.}~\bibnamefont {Zapperi}},\ }\href
  {\doibase 10.1038/ncomms15928} {\bibfield  {journal} {\bibinfo  {journal}
  {Nature Communications}\ }\textbf {\bibinfo {volume} {8}},\ \bibinfo {eid}
  {15928} (\bibinfo {year} {2017})}\BibitemShut {NoStop}%
\bibitem [{\citenamefont {Lin}\ \emph {et~al.}(2014)\citenamefont {Lin},
  \citenamefont {Lerner}, \citenamefont {Rosso},\ and\ \citenamefont
  {Wyart}}]{wyart2014}%
  \BibitemOpen
  \bibfield  {author} {\bibinfo {author} {\bibfnamefont {J.}~\bibnamefont
  {Lin}}, \bibinfo {author} {\bibfnamefont {E.}~\bibnamefont {Lerner}},
  \bibinfo {author} {\bibfnamefont {A.}~\bibnamefont {Rosso}}, \ and\ \bibinfo
  {author} {\bibfnamefont {M.}~\bibnamefont {Wyart}},\ }\href {\doibase
  10.1073/pnas.1406391111} {\bibfield  {journal} {\bibinfo  {journal}
  {Proceedings of the National Academy of Sciences}\ }\textbf {\bibinfo
  {volume} {111}},\ \bibinfo {pages} {14382} (\bibinfo {year}
  {2014})}\BibitemShut {NoStop}%
\bibitem [{\citenamefont {Ozawa}\ \emph {et~al.}(2018)\citenamefont {Ozawa},
  \citenamefont {Berthier}, \citenamefont {Biroli}, \citenamefont {Rosso},\
  and\ \citenamefont {Tarjus}}]{Ozawa2018}%
  \BibitemOpen
  \bibfield  {author} {\bibinfo {author} {\bibfnamefont {M.}~\bibnamefont
  {Ozawa}}, \bibinfo {author} {\bibfnamefont {L.}~\bibnamefont {Berthier}},
  \bibinfo {author} {\bibfnamefont {G.}~\bibnamefont {Biroli}}, \bibinfo
  {author} {\bibfnamefont {A.}~\bibnamefont {Rosso}}, \ and\ \bibinfo {author}
  {\bibfnamefont {G.}~\bibnamefont {Tarjus}},\ }\href {\doibase
  10.1073/pnas.1806156115} {\bibfield  {journal} {\bibinfo  {journal}
  {Proceedings of the National Academy of Sciences}\ }\textbf {\bibinfo
  {volume} {115}},\ \bibinfo {pages} {6656} (\bibinfo {year} {2018})},\ \Eprint
  {http://arxiv.org/abs/https://www.pnas.org/content/115/26/6656.full.pdf}
  {https://www.pnas.org/content/115/26/6656.full.pdf} \BibitemShut {NoStop}%
\bibitem [{\citenamefont {Talamali}\ \emph {et~al.}(2011)\citenamefont
  {Talamali}, \citenamefont {Pet\"aj\"a}, \citenamefont {Vandembroucq},\ and\
  \citenamefont {Roux}}]{Talamali2011}%
  \BibitemOpen
  \bibfield  {author} {\bibinfo {author} {\bibfnamefont {M.}~\bibnamefont
  {Talamali}}, \bibinfo {author} {\bibfnamefont {V.}~\bibnamefont
  {Pet\"aj\"a}}, \bibinfo {author} {\bibfnamefont {D.}~\bibnamefont
  {Vandembroucq}}, \ and\ \bibinfo {author} {\bibfnamefont {S.}~\bibnamefont
  {Roux}},\ }\href {\doibase 10.1103/PhysRevE.84.016115} {\bibfield  {journal}
  {\bibinfo  {journal} {Phys. Rev. E}\ }\textbf {\bibinfo {volume} {84}},\
  \bibinfo {pages} {016115} (\bibinfo {year} {2011})}\BibitemShut {NoStop}%
\bibitem [{\citenamefont {Budrikis}\ and\ \citenamefont
  {Zapperi}(2013)}]{Budrikis2013}%
  \BibitemOpen
  \bibfield  {author} {\bibinfo {author} {\bibfnamefont {Z.}~\bibnamefont
  {Budrikis}}\ and\ \bibinfo {author} {\bibfnamefont {S.}~\bibnamefont
  {Zapperi}},\ }\href {\doibase 10.1103/PhysRevE.88.062403} {\bibfield
  {journal} {\bibinfo  {journal} {Phys. Rev. E}\ }\textbf {\bibinfo {volume}
  {88}},\ \bibinfo {pages} {062403} (\bibinfo {year} {2013})}\BibitemShut
  {NoStop}%
\bibitem [{\citenamefont {Sandfeld}\ \emph {et~al.}(2015)\citenamefont
  {Sandfeld}, \citenamefont {Budrikis}, \citenamefont {Zapperi},\ and\
  \citenamefont {Fernandez~Castellanos}}]{Sandfeld2015}%
  \BibitemOpen
  \bibfield  {author} {\bibinfo {author} {\bibfnamefont {S.}~\bibnamefont
  {Sandfeld}}, \bibinfo {author} {\bibfnamefont {Z.}~\bibnamefont {Budrikis}},
  \bibinfo {author} {\bibfnamefont {S.}~\bibnamefont {Zapperi}}, \ and\
  \bibinfo {author} {\bibfnamefont {D.}~\bibnamefont {Fernandez~Castellanos}},\
  }\href {http://stacks.iop.org/1742-5468/2015/i=2/a=P02011} {\bibfield
  {journal} {\bibinfo  {journal} {Journal of Statistical Mechanics: Theory and
  Experiment}\ }\textbf {\bibinfo {volume} {2015}},\ \bibinfo {pages} {P02011}
  (\bibinfo {year} {2015})}\BibitemShut {NoStop}%
\bibitem [{\citenamefont {T\"uszes}\ \emph {et~al.}(2017)\citenamefont
  {T\"uszes}, \citenamefont {Ispanovity},\ and\ \citenamefont
  {Zaiser}}]{Tuszes2017}%
  \BibitemOpen
  \bibfield  {author} {\bibinfo {author} {\bibfnamefont {D.}~\bibnamefont
  {T\"uszes}}, \bibinfo {author} {\bibfnamefont {P.}~\bibnamefont
  {Ispanovity}}, \ and\ \bibinfo {author} {\bibfnamefont {M.}~\bibnamefont
  {Zaiser}},\ }\href@noop {} {\bibfield  {journal} {\bibinfo  {journal}
  {International Journal of Fracture}\ }\textbf {\bibinfo {volume} {YY}},\
  \bibinfo {pages} {YYY} (\bibinfo {year} {2017})},\ \Eprint
  {http://arxiv.org/abs/arXiv: 1604.01821} {arXiv:arXiv: 1604.01821}
  \BibitemShut {NoStop}%
\bibitem [{\citenamefont {Merabia}\ and\ \citenamefont
  {Detcheverry}(2016)}]{Merabia2017}%
  \BibitemOpen
  \bibfield  {author} {\bibinfo {author} {\bibfnamefont {S.}~\bibnamefont
  {Merabia}}\ and\ \bibinfo {author} {\bibfnamefont {F.}~\bibnamefont
  {Detcheverry}},\ }\href {\doibase 10.1209/0295-5075/116/46003} {\bibfield
  {journal} {\bibinfo  {journal} {EPL}\ }\textbf {\bibinfo {volume} {116}},\
  \bibinfo {pages} {46003} (\bibinfo {year} {2016})}\BibitemShut {NoStop}%
\bibitem [{\citenamefont {Castellanos}\ and\ \citenamefont
  {Zaiser}(2018)}]{Castellanos2018}%
  \BibitemOpen
  \bibfield  {author} {\bibinfo {author} {\bibfnamefont {D.~F.}\ \bibnamefont
  {Castellanos}}\ and\ \bibinfo {author} {\bibfnamefont {M.}~\bibnamefont
  {Zaiser}},\ }\href {\doibase 10.1103/PhysRevLett.121.125501} {\bibfield
  {journal} {\bibinfo  {journal} {Phys. Rev. Lett.}\ }\textbf {\bibinfo
  {volume} {121}},\ \bibinfo {pages} {125501} (\bibinfo {year}
  {2018})}\BibitemShut {NoStop}%
\bibitem [{\citenamefont {Bouttes}\ and\ \citenamefont
  {Vandembroucq}(2013)}]{Bouttes2013}%
  \BibitemOpen
  \bibfield  {author} {\bibinfo {author} {\bibfnamefont {D.}~\bibnamefont
  {Bouttes}}\ and\ \bibinfo {author} {\bibfnamefont {D.}~\bibnamefont
  {Vandembroucq}},\ }\href {\doibase 10.1063/1.4794621} {\bibfield  {journal}
  {\bibinfo  {journal} {AIP Conference Proceedings}\ }\textbf {\bibinfo
  {volume} {1518}},\ \bibinfo {pages} {481} (\bibinfo {year} {2013})},\ \Eprint
  {http://arxiv.org/abs/http://aip.scitation.org/doi/pdf/10.1063/1.4794621}
  {http://aip.scitation.org/doi/pdf/10.1063/1.4794621} \BibitemShut {NoStop}%
\bibitem [{\citenamefont {Liu}\ \emph {et~al.}(2018)\citenamefont {Liu},
  \citenamefont {Ferrero}, \citenamefont {Martens},\ and\ \citenamefont
  {Barrat}}]{Liu2018}%
  \BibitemOpen
  \bibfield  {author} {\bibinfo {author} {\bibfnamefont {C.}~\bibnamefont
  {Liu}}, \bibinfo {author} {\bibfnamefont {E.~E.}\ \bibnamefont {Ferrero}},
  \bibinfo {author} {\bibfnamefont {K.}~\bibnamefont {Martens}}, \ and\
  \bibinfo {author} {\bibfnamefont {J.-L.}\ \bibnamefont {Barrat}},\ }\href
  {\doibase 10.1039/C8SM01392F} {\bibfield  {journal} {\bibinfo  {journal}
  {Soft Matter}\ }\textbf {\bibinfo {volume} {14}},\ \bibinfo {pages} {8306}
  (\bibinfo {year} {2018})}\BibitemShut {NoStop}%
\bibitem [{\citenamefont {Alava}\ \emph {et~al.}(2009)\citenamefont {Alava},
  \citenamefont {Nukala},\ and\ \citenamefont {Zapperi}}]{Alava2009_JPD}%
  \BibitemOpen
  \bibfield  {author} {\bibinfo {author} {\bibfnamefont {M.~J.}\ \bibnamefont
  {Alava}}, \bibinfo {author} {\bibfnamefont {P.~K. V.~V.}\ \bibnamefont
  {Nukala}}, \ and\ \bibinfo {author} {\bibfnamefont {S.}~\bibnamefont
  {Zapperi}},\ }\href {http://stacks.iop.org/0022-3727/42/i=21/a=214012}
  {\bibfield  {journal} {\bibinfo  {journal} {Journal of Physics D: Applied
  Physics}\ }\textbf {\bibinfo {volume} {42}},\ \bibinfo {pages} {214012}
  (\bibinfo {year} {2009})}\BibitemShut {NoStop}%
\bibitem [{\citenamefont {Liu}\ \emph {et~al.}(2016)\citenamefont {Liu},
  \citenamefont {Ferrero}, \citenamefont {Puosi}, \citenamefont {Barrat},\ and\
  \citenamefont {Martens}}]{Liu2016_PRL}%
  \BibitemOpen
  \bibfield  {author} {\bibinfo {author} {\bibfnamefont {C.}~\bibnamefont
  {Liu}}, \bibinfo {author} {\bibfnamefont {E.~E.}\ \bibnamefont {Ferrero}},
  \bibinfo {author} {\bibfnamefont {F.}~\bibnamefont {Puosi}}, \bibinfo
  {author} {\bibfnamefont {J.-L.}\ \bibnamefont {Barrat}}, \ and\ \bibinfo
  {author} {\bibfnamefont {K.}~\bibnamefont {Martens}},\ }\href {\doibase
  10.1103/PhysRevLett.116.065501} {\bibfield  {journal} {\bibinfo  {journal}
  {Phys. Rev. Lett.}\ }\textbf {\bibinfo {volume} {116}},\ \bibinfo {pages}
  {065501} (\bibinfo {year} {2016})}\BibitemShut {NoStop}%
\bibitem [{\citenamefont {Gruber}(1978)}]{Gruber1978}%
  \BibitemOpen
  \bibfield  {author} {\bibinfo {author} {\bibfnamefont {E.}~\bibnamefont
  {Gruber}},\ }\href {\doibase 10.1002/bbpc.19780820975} {\bibfield  {journal}
  {\bibinfo  {journal} {Berichte der Bunsengesellschaft für physikalische
  Chemie}\ }\textbf {\bibinfo {volume} {82}},\ \bibinfo {pages} {1019}
  (\bibinfo {year} {1978})},\ \Eprint
  {http://arxiv.org/abs/https://onlinelibrary.wiley.com/doi/pdf/10.1002/bbpc.19780820975}
  {https://onlinelibrary.wiley.com/doi/pdf/10.1002/bbpc.19780820975}
  \BibitemShut {NoStop}%
\bibitem [{\citenamefont {Zhang}\ \emph {et~al.}(2017)\citenamefont {Zhang},
  \citenamefont {Wang}, \citenamefont {Li}, \citenamefont {Jiang},
  \citenamefont {Li},\ and\ \citenamefont {Liu}}]{Zhang2017}%
  \BibitemOpen
  \bibfield  {author} {\bibinfo {author} {\bibfnamefont {M.}~\bibnamefont
  {Zhang}}, \bibinfo {author} {\bibfnamefont {Y.~M.}\ \bibnamefont {Wang}},
  \bibinfo {author} {\bibfnamefont {F.~X.}\ \bibnamefont {Li}}, \bibinfo
  {author} {\bibfnamefont {S.~Q.}\ \bibnamefont {Jiang}}, \bibinfo {author}
  {\bibfnamefont {M.~Z.}\ \bibnamefont {Li}}, \ and\ \bibinfo {author}
  {\bibfnamefont {L.}~\bibnamefont {Liu}},\ }\href {\doibase
  10.1038/s41598-017-00768-7} {\bibfield  {journal} {\bibinfo  {journal}
  {Scientific Reports}\ }\textbf {\bibinfo {volume} {7}},\ \bibinfo {pages}
  {625} (\bibinfo {year} {2017})}\BibitemShut {NoStop}%
\bibitem [{\citenamefont {Warren}\ and\ \citenamefont
  {Rottler}(2008)}]{Warren2008}%
  \BibitemOpen
  \bibfield  {author} {\bibinfo {author} {\bibfnamefont {M.}~\bibnamefont
  {Warren}}\ and\ \bibinfo {author} {\bibfnamefont {J.}~\bibnamefont
  {Rottler}},\ }\href {\doibase 10.1103/PhysRevE.78.041502} {\bibfield
  {journal} {\bibinfo  {journal} {Phys. Rev. E}\ }\textbf {\bibinfo {volume}
  {78}},\ \bibinfo {pages} {041502} (\bibinfo {year} {2008})}\BibitemShut
  {NoStop}%
\bibitem [{\citenamefont {Schmid}\ and\ \citenamefont
  {Grasso}(2012)}]{Schmid2012}%
  \BibitemOpen
  \bibfield  {author} {\bibinfo {author} {\bibfnamefont {A.}~\bibnamefont
  {Schmid}}\ and\ \bibinfo {author} {\bibfnamefont {J.-R.}\ \bibnamefont
  {Grasso}},\ }\href {\doibase 10.1029/2011JB008975} {\bibfield  {journal}
  {\bibinfo  {journal} {Journal of Geophysical Research: Solid Earth}\ }\textbf
  {\bibinfo {volume} {117}},\ \bibinfo {pages} {n/a} (\bibinfo {year}
  {2012})},\ \bibinfo {note} {b07302}\BibitemShut {NoStop}%
\bibitem [{\citenamefont {Popovi\ifmmode~\acute{c}\else \'{c}\fi{}}\ \emph
  {et~al.}(2018)\citenamefont {Popovi\ifmmode~\acute{c}\else \'{c}\fi{}},
  \citenamefont {de~Geus},\ and\ \citenamefont {Wyart}}]{wyart2018}%
  \BibitemOpen
  \bibfield  {author} {\bibinfo {author} {\bibfnamefont {M.}~\bibnamefont
  {Popovi\ifmmode~\acute{c}\else \'{c}\fi{}}}, \bibinfo {author} {\bibfnamefont
  {T.~W.~J.}\ \bibnamefont {de~Geus}}, \ and\ \bibinfo {author} {\bibfnamefont
  {M.}~\bibnamefont {Wyart}},\ }\href {\doibase 10.1103/PhysRevE.98.040901}
  {\bibfield  {journal} {\bibinfo  {journal} {Phys. Rev. E}\ }\textbf {\bibinfo
  {volume} {98}},\ \bibinfo {pages} {040901} (\bibinfo {year}
  {2018})}\BibitemShut {NoStop}%
\end{thebibliography}%

\end{document}